\documentclass[12pt]{article}
\usepackage{blindtext}
\usepackage[a4paper, total={6.5in, 10in}]{geometry}
\usepackage{graphicx}
\usepackage{subfig}
\usepackage{xcolor}
\usepackage{lineno}
\usepackage{multirow}
\usepackage{soul}
\usepackage{float}
\usepackage{soul,color}
\usepackage{hyperref}
\usepackage{amsmath}
\usepackage{amssymb}
\usepackage{booktabs}
\usepackage{authblk}
\usepackage{textcomp}
\usepackage{caption}
\usepackage{lipsum}
\usepackage{comment}
\usepackage{lineno}
\usepackage[utf8]{inputenc}
\providecommand{\keywords}[1]
{
  \small	
  \textbf{\textit{Keywords---}} #1
}

\title{Collective Intelligent Strategy for Improved Segmentation of COVID-19 from CT}

\author{Surochita Pal Das\footnote{corresponding Author. \\ email(s): pal.surochita@gmail.com;  sushmita@isical.ac.in; uma@isical.ac.in \\ orcid(s): 0000-0002-6749-4426 (S. P. Das), 0000-0001-9285-1117 (S. Mitra), 0000-0003-0332-2294 (B. Uma Shankar)\\ All the authors  contributed equally to this research.\\ This work was supported by the J. C. Bose National Fellowship, sanction no. JCB/2020/000033 of S. Mitra} }

\author{Sushmita Mitra}
\author{B. Uma Shankar}

\affil{Machine Intelligence Unit, Indian Statistical Institute, 203, B.T. Road, Kolkata, 700108, West Bengal, India.}

\begin{document}
\maketitle

\abstract{The devastation caused by the coronavirus pandemic makes it imperative to design automated techniques for a fast and accurate detection. We propose a novel non-invasive tool, using deep learning and imaging, for delineating COVID-19 infection in lungs. The Ensembling Attention-based Multi-scaled Convolution network (EAMC), employing Leave-One-Patient-Out (LOPO) training, exhibits high sensitivity and precision in outlining infected regions along with assessment of severity. The  Attention module combines contextual with local information, at multiple scales, for accurate segmentation. Ensemble learning integrates heterogeneity of decision through different base classifiers. The  superiority of EAMC, even with severe class imbalance, is established through comparison  with existing state-of-the-art learning models over four publicly-available COVID-19 datasets. The results are suggestive of the relevance of deep learning in providing assistive intelligence to medical practitioners, when they are overburdened with patients as in pandemics. Its clinical significance lies in its unprecedented scope in providing low-cost decision-making for patients lacking specialized healthcare at remote locations.}


\section{Introduction}

The recent pandemic, called the novel coronavirus-disease-2019 (COVID-19), has been a major threat to world-health ~\cite{siddiqi2021covid}; with medical systems collapsing around the globe. It resulted in an increasing demand for health services, encompassing finite components like beds, critical medical equipment, and healthcare workers (who also get regularly infected). Even the year 2022 has seen proliferation of  newer strains of the virus affecting humankind.  Some of the major COVID-19 complications, in case of serious level of infection,  include acute respiratory distress syndrome (ARDS), pneumonia, multi-organ failure, septic-shock, and even death. Serious illness is more likely to result in people with existing co-morbidities. Often there exist long term side-effects in post-COVID patients.

An early detection, diagnosis, isolation and prognosis, play a major role in controlling the spread of the disease. Computed tomography (CT) and X-rays are the commonly used imaging techniques for the lung. The CT scan uses X-rays to produce a 3D view comprising cross-sectional slices, for detecting existing anomalies.
Occurrence of false negatives in the ``gold standard" RT-PCR test results often lead to the chest CT scans being an useful supplement in projecting typical infection characteristics -- like Ground-Glass Opacity (GGO) and/or mixed  consolidations.
It was reported \cite{fang2020sensitivity} that Lung CT images are more sensitive (98\%), as compared to RT-PCR (71\%), in correctly predicting COVID-19.

Doctors reported difference in CT abnormalities related to COVID-19 patients in multiple studies~\cite{ding2020chest,li2020coronavirus}.
It was observed, even at early stages, that viral infections were indicated by clear patterns~\cite{barstugan2020coronavirus,ding2020chest}.
In Ref. \cite{li2020coronavirus} the researchers  assessed the effectiveness of chest CT in the diagnosis and treatment of COVID-19. The CT characteristics of COVID-19 were presented and compared with the manifestations of other viruses.

Abnormalities in CT may occur~\cite{ding2020chest} before the appearance of clinical symptoms. Multifocal, unilateral, and peripherally based GGO are examples of classic patterns, which are also observed in symptomatic cases. Abnormalities like inter-lobular septal thickening, thickening of the surrounding pleura, round cystic alterations, nodules, pleural effusion, bronchiectasis, and lymphadenopathy were infrequently detected in the asymptomatic group.

Manually detecting COVID-affected regions from lung CT scans is time consuming and prone to inherent human bias. Thus automated or semi-automated Computer-Aided-Diagnosis (CAD) becomes necessary \cite{BANERJEE2018337, parmar2014robust}. An accurate, automated detection and delineation of the COVID-19 infection is of great importance since this results in an effective monitoring of its spread within the lungs. This helps in predicting the severity of the infection, as well as its prognosis.

Smart machines can imitate the human brain to some extent. Everything that makes a machine smarter falls under the umbrella called Artificial Intelligence (AI). Machine Learning (ML), which is a subset of AI, consists of a collection of algorithms and tools which enable a machine to understand patterns within the data without being explicitly programmed. ML uses this underlying structure to perform logical reasoning for a task. Deep Learning (DL), again, is a sub-domain of ML~\cite{Goodfellow2016}.
 It aids a machine in learning hidden patterns within the data without any expert intervention, to make predictions -- given high computational power and a massive volume of annotated data.  A convolution neural network (CNN), which is a DL model, has been shown to perform effectively in analyzing visual images. A CNN  model, which was designed to recognize objects in natural-images from the ImageNet Large-Scale Visual Recognition Challenge (ILSVRC), was found to be comparable in efficiency to humans~\cite{krizhevsky2017imagenet}.

The $U$-Net~\cite{ronneberger2015u} is an encoder-decoder type of CNN architecture, designed for the segmentation of biomedical images in a fast and precise manner. The encoder arm causes the spatial dimension to be decreased, while increasing the number of channels. In contrast, the decoder arm decreases the channels while raising the spatial dimensions. Introduction of attention gates (AGs)~\cite{oktay2018attention} in the $U$-Net framework, help reduce the feature responses in irrelevant background regions, while providing more weight to region of interest (ROI). The network is guided towards learning only the relevant information in terms of the weighted local features. Incorporating dilated convolutions~\cite{geng2019lung} allows feature extraction at multiple scales.

 Multi-scalar approaches, which observe and evaluate a dataset at several scales, are popular in the machine learning domain. They capture the local geometry of neighbourhoods, which are characterised by a collection of distances between points or groups of closest neighbours. This is analogous to looking at a portion of a slide at various microscopic resolutions; whereby, very small features can be detected at high resolution from a restricted region of the sample. As the majority portion of the slide is examined at a lower resolution, it allows one to examine the larger (global) features as well. Multi-scalar methods have been found to perform better than state-of-the-art techniques, with reduced sample sizes, in the medical domain ~\cite{wang2018interactive}.

An ensemble-based classifier system is designed by merging multiple diverse classifier models.  Ensembling makes statistical sense in a number of situations. We regularly employ such an approach in our daily lives while seeking the advice of various experts prior to taking a major decision.
For instance, we frequently seek the advice of numerous doctors before consenting to a medical procedure.
The main objective is to reduce the regrettable choice of a needless medical procedure.
The experts must differ from one another in some way for this mechanism to be successful. Individual classifiers, due to their inherent diversity, can produce various decision limits within the context of classification. This is commonly achieved by employing distinct training setup for each classifier.
If adequate diversity is established, each classifier will commit a separate error, which may then be strategically combined to lower the overall error~\cite{rabi2006}.

Advantage of ensemble learning lies in its inherent diversity. This can be introduced by embedding different training datasets, or  features, or classifiers; or even differing initialization and/or parameters of the classifier(s) involved.
According to Dietterich~\cite{dietterich2000ensemble} there are three main justifications for employing an ensemble-based system, {\it viz.}
statistical, computational, and representational.
The computational criterion refers to the model selection problem.
The statistical cause is connected to the insufficiency of available data to accurately represent a distribution.
The representational cause addresses situations where the selected model is unable to accurately represent the desired decision boundary.

Numerous studies have been reported in recent times in the domain of COVID-19, using neural networks and data-driven algorithms. These include machine learning approaches for diagnosis of COVID-19 from X-Ray/ CT images~\cite{oh2020deep,roberts2021common,wang2020fully}. A pre-trained deep-learning model, called  DenseNet, was developed~\cite{wang2020fully} for classifying 121  CT-images into COVID-19 positive and negative categories. Application of the ResNet-18 was made~\cite{zhang2020clinically} to segment and classify lung-lesions of COVID-19, pneumonia infection, and normal ones.

A deep learning based AI system was designed  \cite{zhou2022interpretable} to detect and quantify lesions from chest CT. It can remove scan-level bias to extract precise radiomic features.
The Unified CT-COVID AI Diagnostic Initiative (UCADI)~\cite{bai2021advancing} enables independent training at each host institution, under a combined learning framework, without data sharing. This was shown to outperform the local models, thereby advancing the prospects of utilizing combined learning for privacy-preserving AI in digital health.

Deep learning has been employed for evaluating the severity of COVID-19 infection~\cite{wu2020severity}.
Well-known deep models, like $U$-Net~\cite{ronneberger2015u}, Residual $U$-Net \cite{resunt-2018}, Attention $U$-Net~\cite{oktay2018attention}, have been used for screening COVID-19. There exist ensemble methods for segmentation of CT images~\cite{ben2022Ensemble,Enshaei2021Ensemble}. The Inception-V3, Xception, InceptionResNet-V2 and DenseNet-121 were ensembled \cite{Enshaei2021Ensemble} for a multiclass segmentation of GGO and Consolidation in COVID-19 CT data over the data CT-Seg (Table~\ref{data}). Each of these models used the CNN as backbone, with pre-trained weights from ImageNet being further trained over the CT-Seg data. The split into training, validation and  testing sets were 40, 10, 50 images, respectively, with pixel-level soft majority voting being employed for their aggregation.

A cascade of two $U$-Nets, with VGG backbone, was ensembled~\cite{ben2022Ensemble} to extract the lung region, followed by the delineation of the GGO and consolidation regions. Multiclass segmentation of GGO and Consolidation was performed over CT-Seg, Seg-nr.2 and Kaggle-COVID-19  datasets (Table~\ref{data}) along with some private dataset; while the training data contained parts of CT-Seg and Seg-nr.2, the remaining parts of the datasets were used for testing the model. The training process of each network in the ensemble differed due to random weight initialization, and data augmentation with shuffling.

Domain Extension Transfer Learning (DETL) was employed \cite{sanhita2020} for the screening of COVID-19, with characteristic features being determined from chest X-Ray images. In order to get an idea about the COVID-19 detection transparency, the authors  employed the concept of Gradient Class Activation Map (Grad-CAM) for detecting the regions where the model paid more attention during the classification. The results are claimed to be strongly correlated with clinical findings.

\begin{figure}[!ht]
    \centering
    \includegraphics[width=15cm]{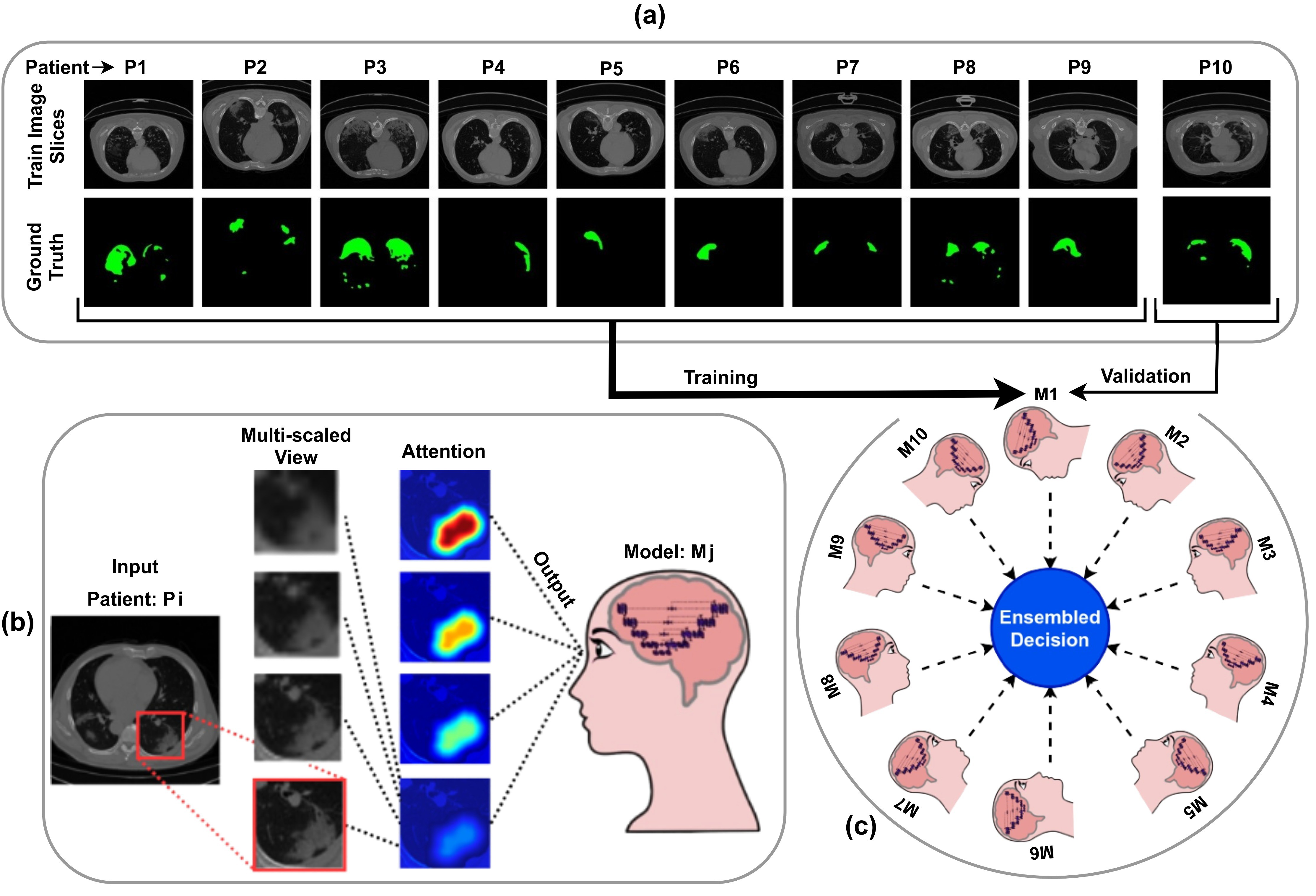}
    \caption{Schematic representation of Ensembling with LOPO  the Attention-based Multiscaled CNNs. (a) Leave-One-Patient-Out scheme $LOPO$. (b) Attention-based Multiscaled view in $AMC$. (c) Ensembling of the $AMC$ Models}
    \label{fig:visual}
\end{figure}

\section{Results}\label{Results}

We propose a novel Ensembled Attention-based Multi-scaled Convolution networks ($EAMC$), using $LOPO$ learning, based on CT images of TEN patients  and trained using TEN base-classifiers. As each classifier takes only nine samples (patients) for training, and starts from scratch, it does result in a completely new classifier with different set of parameters. The remaining ONE sample is left for validation in each case (as elaborated in Section~\ref{implement}). These TEN trained classifiers are ensembled to segment the COVID-infection region (ROI)  through majority voting, over four different test datasets collected from various publicly available sources (Table~\ref{data}). The workflow of the $EAMC$ is visualized in Fig. \ref{fig:visual}.

\begin{figure}[!ht]
    \centering
    \includegraphics[width=15cm]{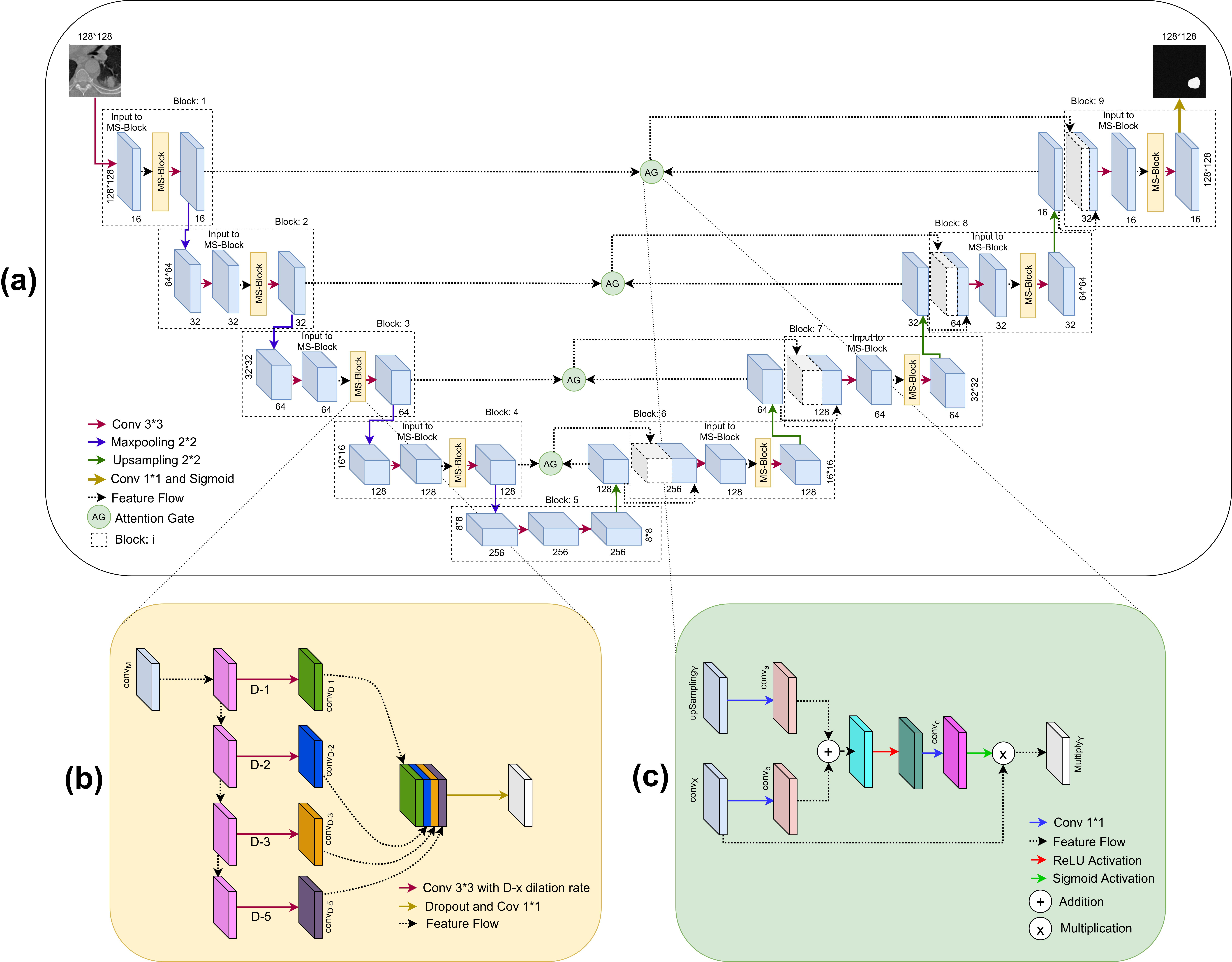}
    \caption{(a) The {\it AMC-Net} framework, with detailed representation of (b) of $MS$-block, and (c) Attention Gate ($AG$)}
    \label{fig:AMCmodel}
\end{figure}

The objective of the model is to segment a COVID-19 infected CT lung image into its Regions of Interest (ROI) [{\it i.e.}, GGO and Consolidation], and background (containing all other regions in the image). This is not an easy task for  a vanilla $U$-Net. Therefore, an $AG$ is carefully incorporated to focus  on the ROI,  whereas the multi-scalar  dilation provides the necessary local and neighbourhood information representation for delineating the ROI.  Focal Loss ($FL$) [eqn. (\ref{focalLoss})] is used as the loss function to compensate for class imbalance. Finally we implement ensembling by $LOPO$ for training with limited annotated data. The details about the training dataset are provided in Section \ref{datasets}.  Validation is successfully performed on unseen, independently-compiled data from multiple publicly-available data sources. The proposed $EAMC$ of Fig.~\ref{fig:visual}, using LOPO ensemble learning on {\it AMC-Net} of Fig.~\ref{fig:AMCmodel}, demonstrates  good generalization capability. It is efficient, accurate, consistent and robust on the unseen data. The architecture of {\it AMC-Net} is summarized in Tables~\ref{AMC-Net Architecture}-\ref{AG-block}.

\subsection{Data used}\label{dataused}

\begin{table}[!ht]
\caption{Breakup of Infected \& Non-Infected samples  and  slices, in the training and test  datasets}
\centering
\scalebox{1}{
\begin{tabular}{@{}llllll@{}}
   \toprule \toprule
    Sr.               & Dataset              &    Total patients &   \begin{tabular}{@{}l@{}}Infected \\ slices\end{tabular} &  \begin{tabular}{@{}l@{}}Non-Infected \\ slices\end{tabular} & Comment \\[0.5ex]
    \midrule
    \multirow{4}{*}{1}    & \begin{tabular}{@{}l@{}}Kaggle-COVID-19$^{a}$: \\ Part-1\end{tabular} &      10          &    1351            & 1230 &  \begin{tabular}{@{}l@{}}Training \\ \end{tabular} \\
    \cline{2-6}
                          & \begin{tabular}{@{}l@{}}Kaggle-COVID-19$^{a}$: \\ Part-2\end{tabular} &       10         &    493            & 446  &   \begin{tabular}{@{}l@{}}Testing \\ \end{tabular} \\ \midrule
    2       & CT-Seg$^b$    &     $>$40         &    100            & 0   &        Testing        \\\midrule
    3       & Seg-nr.2$^b$    &        9          &    372            & 457  &       Testing       \\\midrule
    4       & MosMed$^d$    &        47$^{\S}$          &    761            & 1166  &          Testing      \\\midrule

    \bottomrule
\end{tabular}
}
\label{data}
\raggedright
$^a$Kaggle-COVID-19: {\url{https://www.kaggle.com/datasets/andrewmvd/covid19-ct-scans}~\cite{ma2021toward}}\\
$^b$CT-Seg \&  Seg-nr.2 (Two Datasets):  {\url{http://medicalsegmentation.com/covid19/}}\\
$^c$MosMed: {\url{https://mosmed.ai/en/}}~\cite{morozov2020mosmeddata} \\
$^{\S}$Relevant 47 samples are used (actual available samples being 50).
\end{table}

Four  datasets, with details as provided in  Table~\ref{data}, were used in this study.  The Kaggle-COVID-19 data  comprises  20 patient samples, of which ten (having at least 200 but not more than 301 slices) were kept for training. This was named as Kaggle-COVID-19:Part-1 dataset. The remaining ten samples, each containing $<$ 200 or $>$ 301 slices, were retained for testing. This was termed the Kaggle-COVID-19:Part-2 dataset. The data is  available on the Kaggle platform\footnote{\url{https://www.kaggle.com/datasets/andrewmvd/covid19-ct-scans}} in annotated form  \cite{ma2021toward}.   The other three datasets, of Table~\ref{data}, were also clubbed together for testing. The unseen test datasets were thus used only for evaluating the generalization performance of $EAMC$.

\subsection{Implementational details}\label{implement}

\begin{table}[!ht]
        \centering
        \caption{Description of Ensemble models, with $DSC$ on corresponding validation sets}
        \scalebox{0.78}{
        \begin{tabular}{@{}lllllllllll@{}}
            \toprule
             Model No.& M1 & M2 & M3 & M4 & M5 & M6 & M7 & M8 & M9 & M10 \\
             \midrule
             Training
             & P/\{P10\}
             & P/\{P9\}
             & P/\{P8\}
             & P/\{P7\}
             & P/\{P6\}
             & P/\{P5\}
             & P/\{P4\}
             & P/\{P3\}
             & P/\{P2\}
             & P/\{P1\} \\ \midrule
             Validation & P10 & P9 & P8 & P7 & P6 & P5 & P4 & P3 & P2 & P1 \\ \midrule
             DSC & 0.8882 & 0.8952 & 0.8617 & 0.8814 & 0.8432 & 0.8867 & 0.8944 & 0.8821 & 0.8787 & 0.8615 \\ \midrule
             \multicolumn{11}{l@{}}{ where P = \{ Pi: $i\in\mathbb{N} \wedge i\in[1,10]$ \} }\\\midrule
             \multicolumn{11}{l@{}}{Mean validation DSC = 0.8773, with Standard deviation = 0.0158}\\
             \bottomrule
        \end{tabular}
        }
        \label{tab:ensemble}
    \end{table}

The ten classifier models considered are $M1$, $M2$, $\ldots$, $M10$; with the validation datasets defined as $P10$, $P9$, $\ldots$, $P1$, and described in Table~\ref{tab:ensemble}. In each case, the rest of the corresponding patient's dataset is used for training by the $AMC$-Net model of Fig. \ref{fig:AMCmodel}. For example, in case-1, $M1$ is trained with $P1$ to $P9$ patient datasets and validated on $P10$ patient dataset. Similarly in case-2, $M2$ is trained with $P1$ to $P8$ and $P10$ patient datasets and validated on $P9$ patient dataset, and so on for all the models $M3$ to $M10$. Each model,  $M1$ to $M10$, is trained with different  parameters, pertaining to initialization, learning rate, dropout probability, etc. The batch size was kept uniform at 16, using the Adam optimizer \cite{kingma2014adam} over 70 epochs. Values of learning rate and dropout probability were set at 0.001 and 0.2, respectively, after several experiments. Run time augmentation (rotaion ± 10\textdegree; horizontal shift Range ± 0.2; vertical shift Range ± 0.2; zoom range ± 0.2) was employed during training.

The implementation was made in the Tensorflow framework, running behind wrapper library Keras using python version 3.6, Keras version 2.2.4, and Tensorflow-GPU version 1.13.1, with dedicated GPU (NVIDIA TESLA P6 having capacity of 16GB).

\subsection{Experimental outputs}\label{experiment}

\begin{figure}[!ht]
    \centering
    \includegraphics[width=16cm]{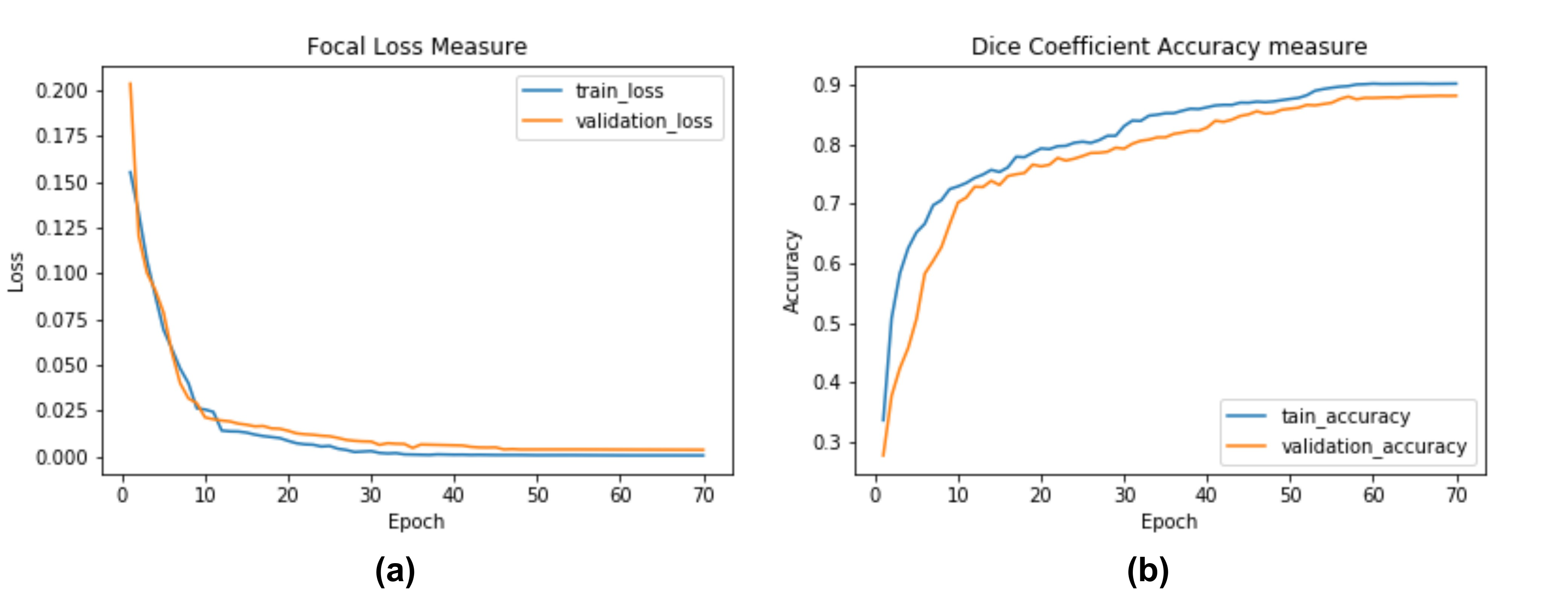}
    \caption{Sample learning curves, for Model M3, over training and validation sets; (a) loss: $FL$, and (b) accuracy: $DSC$}
    \label{fig:LossAccuracy1}
\end{figure}

Sample  learning curves, depicting loss [\textit{Focal Loss (FL)} of eqn. (\ref{focalLoss})] and accuracy [\textit{Dice Score Coefficient (DSC)} of eqn. (\ref{Metric:DiceScore})], are illustrated in Fig.~\ref{fig:LossAccuracy1}. The validation set corresponding to each model, with the resultant  $(DSC)$, are presented in Table~\ref{tab:ensemble}.
The accuracy and robustness of output in each case is evident, showing an average $DSC$ of 0.8773 with a standard deviation (SD) of 0.0158 on the validation datasets.

\begin{figure}[!ht]
    \centering
    \includegraphics[width=15cm]{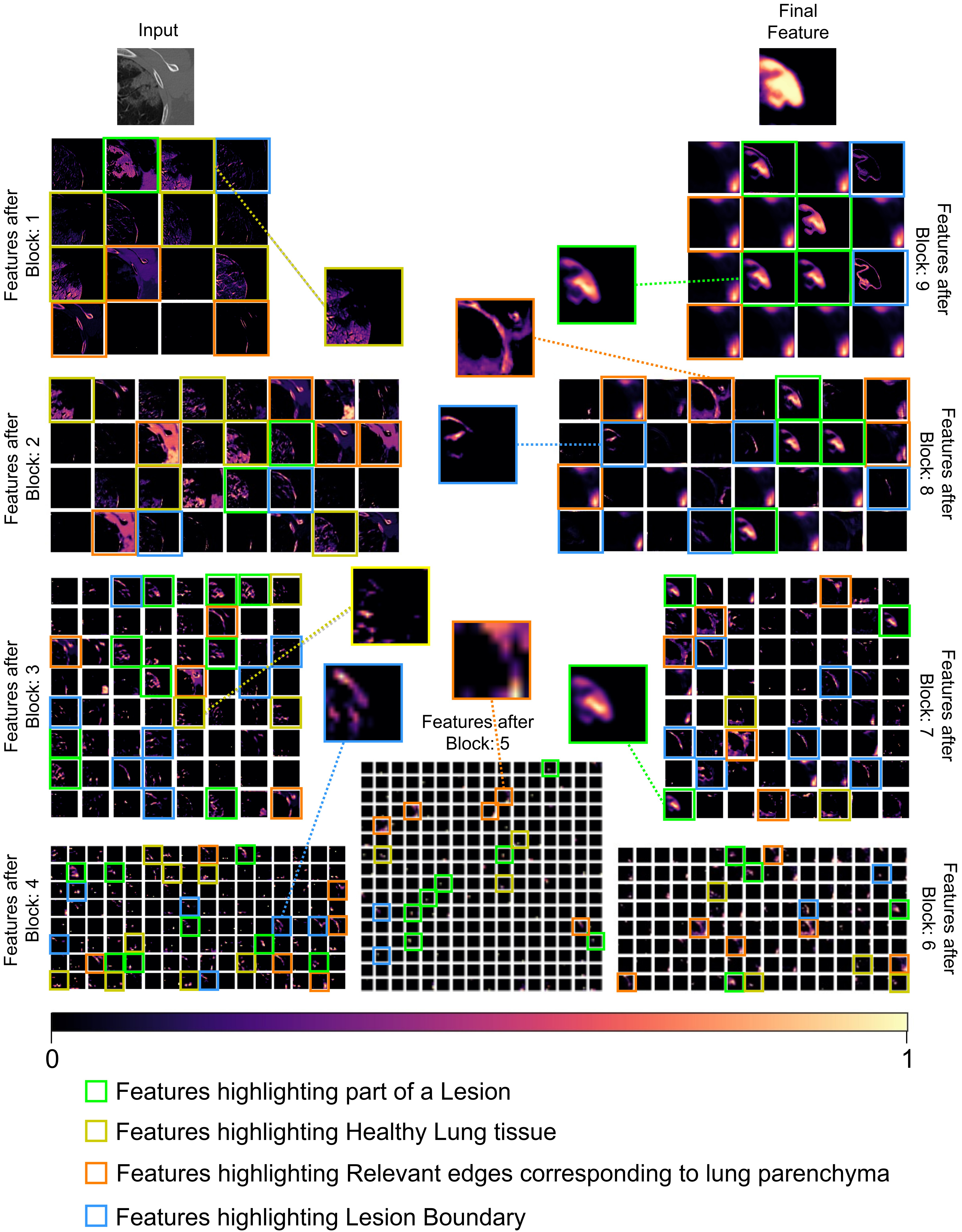}
    \caption{Visualization of features extracted after each Block of the {\it AMC-Net} from Fig \ref{fig:AMCmodel}}
    \label{fig:feature}
\end{figure}

Next the explainability of the {\it AMC-Net} architecture was analysed. As evident from Fig. \ref{fig:feature}, each level of the encoder and decoder arms of the network architecture demonstrate extraction of meaningful features at different levels of abstraction. The abstraction level of the extracted features are dependent on the level of the Block. While some bright objects do get highlighted at the initial stages, as the Block depth increases the Attention and Multi-Scalar mechanisms of the $AMC-Net$ assist in highlighting the ROIs (like, GGO and Consolidation) present in the CT patches and suppress the background.

For example, after Block 1 there are some features highlighting healthy lung tissue (yellow box) in the figure. There are also relevant edges corresponding to lung parenchyma (orange box).  As the Block level increases, the attention over the lesion boundary (blue box) and lesion (green box) in Fig. \ref{fig:feature} become more prominent. Eventually the final feature which prominently emerges is the lesion.

\subsection{Comparative study}\label{compare}

Using the {\it AMC-Net} as the base classifier, with ten classifiers $M1$ to $M10$ ensembled by $LOPO$, results were generated on the four test datasets, {\it viz.} Kaggle-COVID-19: Part2, CT-seg., Seg-nr.2, and MosMed (as described in  Table~\ref{data} and Section~\ref{datasets}) in terms of the performance metrics defined in eqns. (\ref{Metric:DiceScore})-(\ref{Metric:Sensitivity}).

The comparative analysis of output generated by the $EAMC$, on different test sets, is presented in Fig.~\ref{fig:boxplotComparison}. In all cases the test slices were divided into  non-overlapping patches (as elaborated in Section \ref{TrainPatch}). The segmentation output is aggregated using majority voting.

 It is observed that all the metrics provided a consistently better performance over the MosMed data.  This is perhaps because the average intensity of the CT scans in MosMed data is higher than that of the rest of the CT datasets used; thereby, providing a better contrast between the ROI and background. Moreover the Hounsfield range is not distorted here, such that there is less noise present.

 In order to explore the effectiveness of $AMC$-Net as the base classifier in $EAMC$, we also compared four state-of-the-art models like R2UNet \cite{alom2018recurrent}, Inception Net \cite{Enshaei2021Ensemble}, Xception Net \cite{Enshaei2021Ensemble}, and DenseNet \cite{Enshaei2021Ensemble} (which have been extensively employed in COVID-19 segmentation literature). Ensembling of classifiers, in each case, was by $LOPO$ during training  (involving same hyper-parameters). Testing was performed  on the four datasets, as elaborated earlier. A comparative study of the evaluation metrics [eqns. (\ref{Metric:DiceScore})-(\ref{Metric:Sensitivity})] is provided in Fig.~\ref{fig:boxplotComparison}. It is found that all the metrics resulted in higher values for the ensembled {\it AMC-Net}, indicating a better generalization in segmentation over each test dataset.

\begin{figure}[!ht]
  \centering
  \begin{tabular}{@{}c@{}}
    \includegraphics[width=1\linewidth]{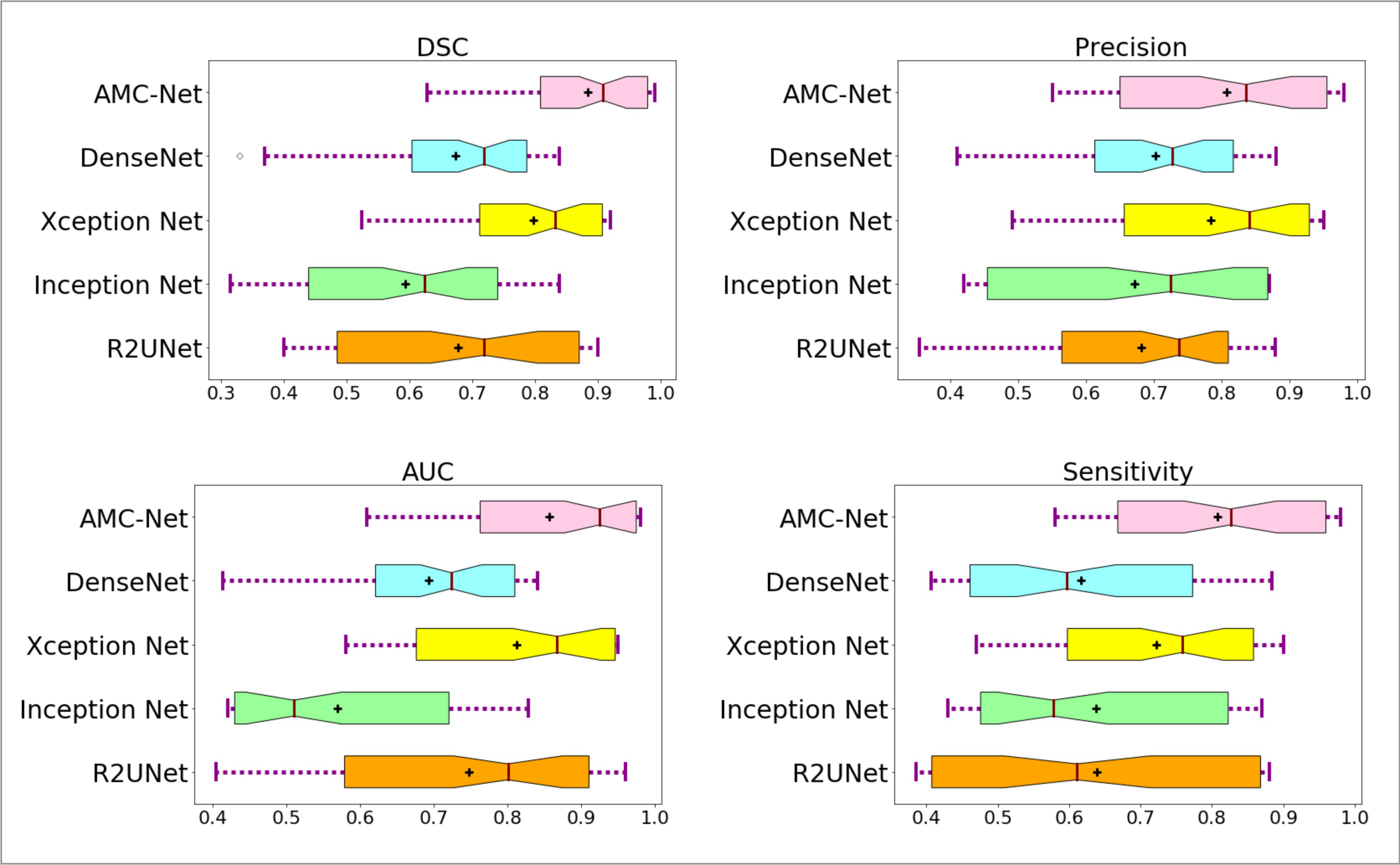} \\
    (a) Comparison over Kaggle Data
  \end{tabular}

  \vspace{\floatsep}

  \begin{tabular}{@{}c@{}}
    \includegraphics[width=1\linewidth]{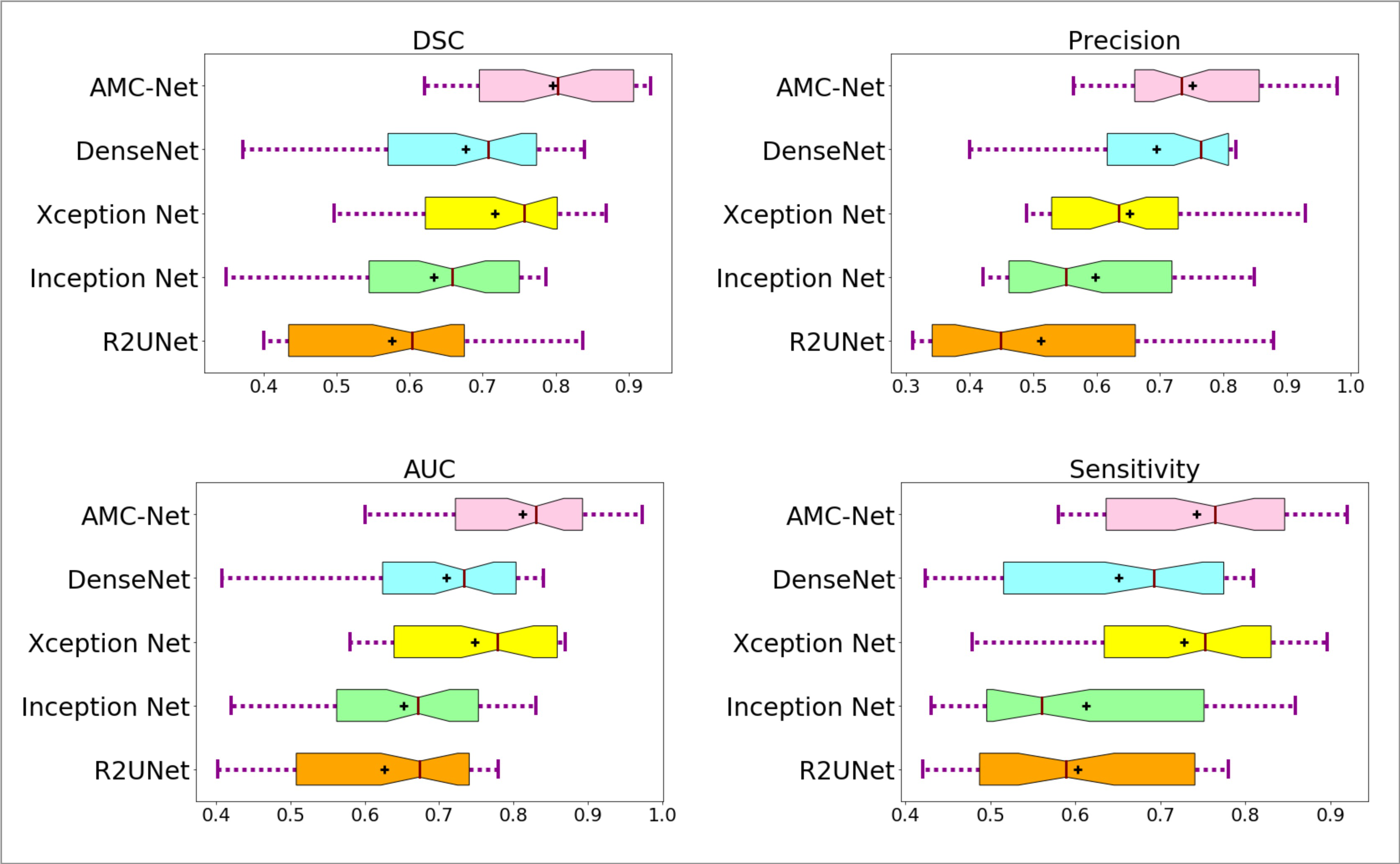} \\
    (b) Comparison over CT-Seg Data
  \end{tabular}

  \caption{Comparative performance evaluation of base classifiers, in the uniform framework of ensembling with $LOPO$, over the test datasets (continued)}
\end{figure}%

\begin{figure}[!ht]\ContinuedFloat
    \centering
  \begin{tabular}{@{}c@{}}
    \includegraphics[width=1\linewidth]{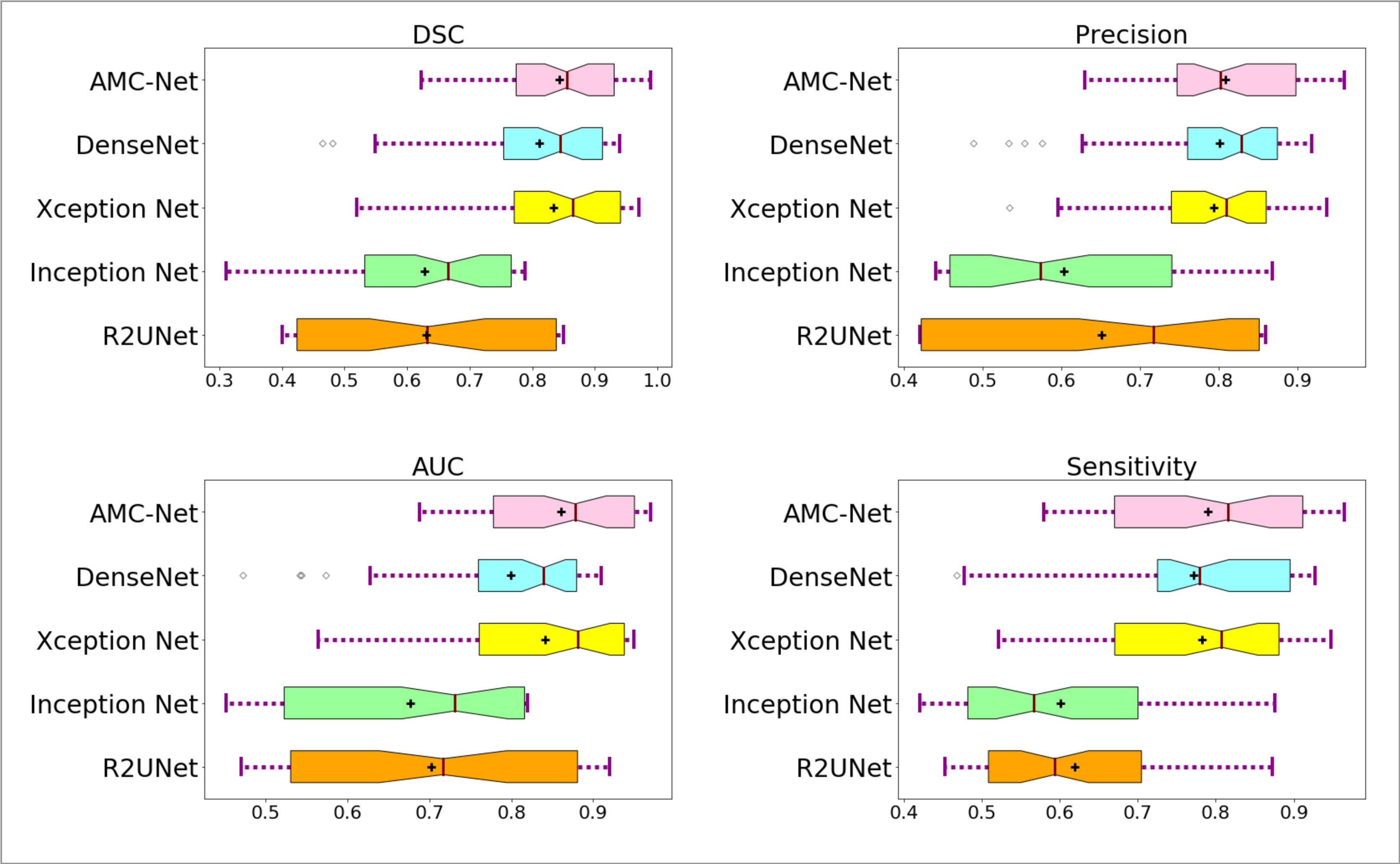} \\
    (c) Comparison over Seg-nr.2 Data
  \end{tabular}

  \vspace{\floatsep}

  \begin{tabular}{@{}c@{}}
    \includegraphics[width=1\linewidth]{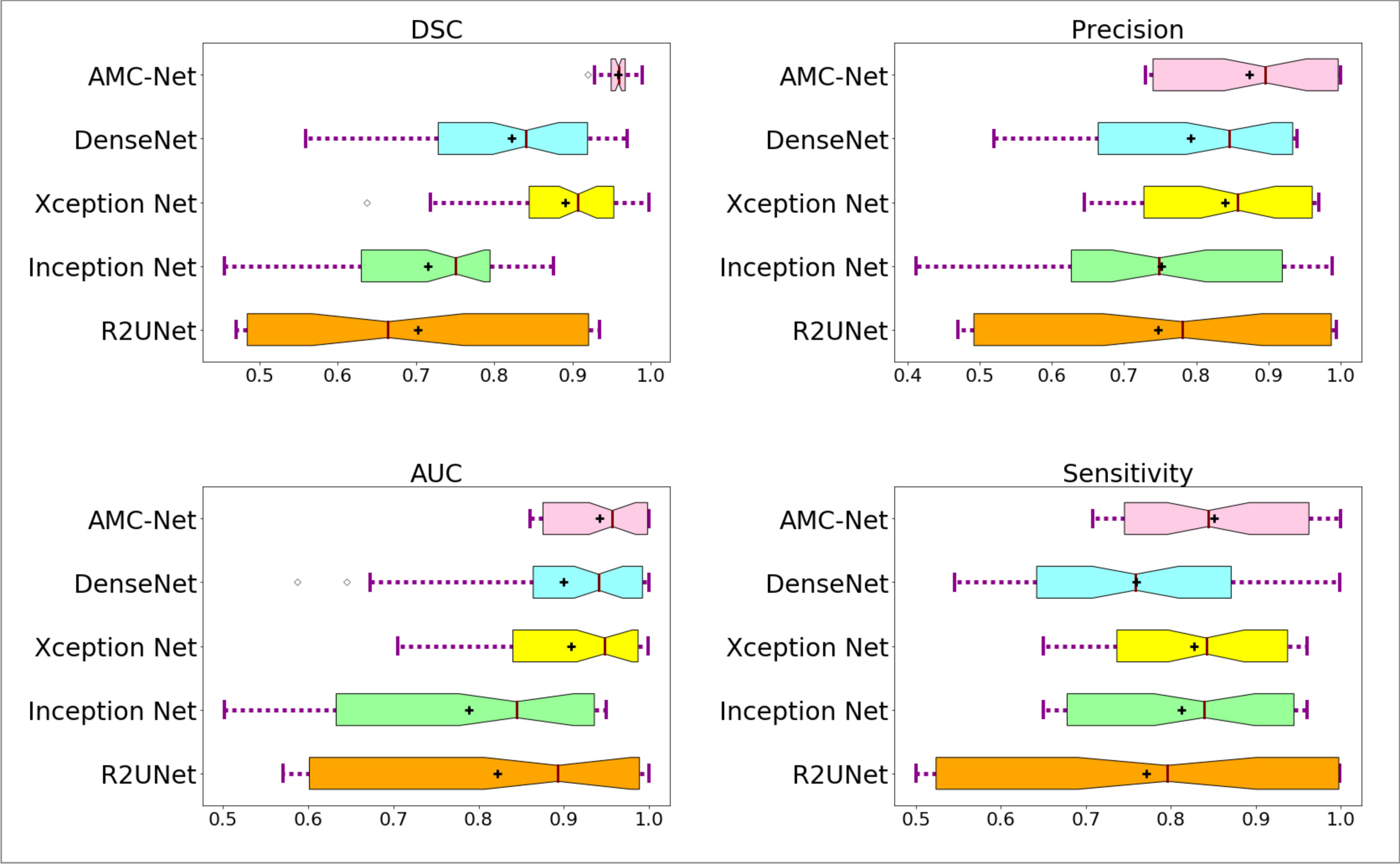} \\
    (d) Comparison over MosMed Data
  \end{tabular}

  \caption{Comparative performance evaluation of base classifiers, in the uniform framework of ensembling with $LOPO$, over the test datasets}
  \label{fig:boxplotComparison}
\end{figure}

Summarized below is the performance of our $EAMC$ over the four test datasets under consideration.
 \begin{itemize} \item Kaggle-COVID-19: $DSC$ 0.8840 ± 0.103, $Precision$ 0.8072 ± 0.148, $AUC$ 0.8562 ± 0.126, $Sensitivity$ 0.8082 ± 0.145; \item CT-Seg: $DSC$ 0.7955 ± 0.107, $Precision$ 0.7500 ± 0.121, $AUC$ 0.8123 ± 0.099, $Sensitivity$ 0.7431 ± 0.114; \item Seg-nr.2: $DSC$ 0.8431 ± 0.101, $Precision$ 0.8089 ± 0.096, $AUC$ 0.8613 ± 0.091, $Sensitivity$ 0.7905 ± 0.125; \item MosMed: $DSC$ 0.9584 ± 0.014, $Precision$ 0.8733 ± 0.112, $AUC$ 0.9417 ± 0.056, $Sensitivity$ 0.8514 ± 0.107. \end{itemize}

An investigation into related results on COVID-19 segmentation with deep networks, as reported in literature \cite{ben2022Ensemble,cong2022boundary,ma2021toward,shabani2022self,sun2022weakly,wang2022regularization} using some of the same test datasets, led to interesting conclusions with respect to our $EAMC$.
The Seg-nr.2 test set yielded $DSC$ of 0.673 \cite{ma2021toward}, 0.620 \cite{shabani2022self}. The model \cite{shabani2022self} employed generative adversarial network (GAN) with $U$-Net as backbone, and reported a $Sensitivity$ of 0.672. Their results on the MosMed test set show $DSC$ 0.584, with $Sensitivity$ 0.768 . Using the Kaggle-COVID-19 dataset the authors reported \cite{wang2022regularization} $DSC$ 0.7103, $Sensitivity$ 0.6860. Combination of the test sets Kaggle-COVID-19 and Seg-nr.2 was also reported. The authors in Ref. \cite{cong2022boundary}, using the R2UNet as backbone, obtained a $DSC$ of 0.851. On the other hand, using the basic $U$-Net as backbone \cite{ben2022Ensemble} attained $DSC$ 0.80.  The Medseg (combination of CT-Seg and Seg-nr.2) dataset resulted in $DSC$  0.77 \cite{sun2022weakly}.
This establishes the effectiveness of our $EAMC$, the ensembled classifier using $AMC$-Net, in terms of these compared performance metrics over all test datasets.

\subsection{Ablations} \label{ablation}

\begin{figure}[!ht]
    \centering
    \includegraphics[width=15cm]{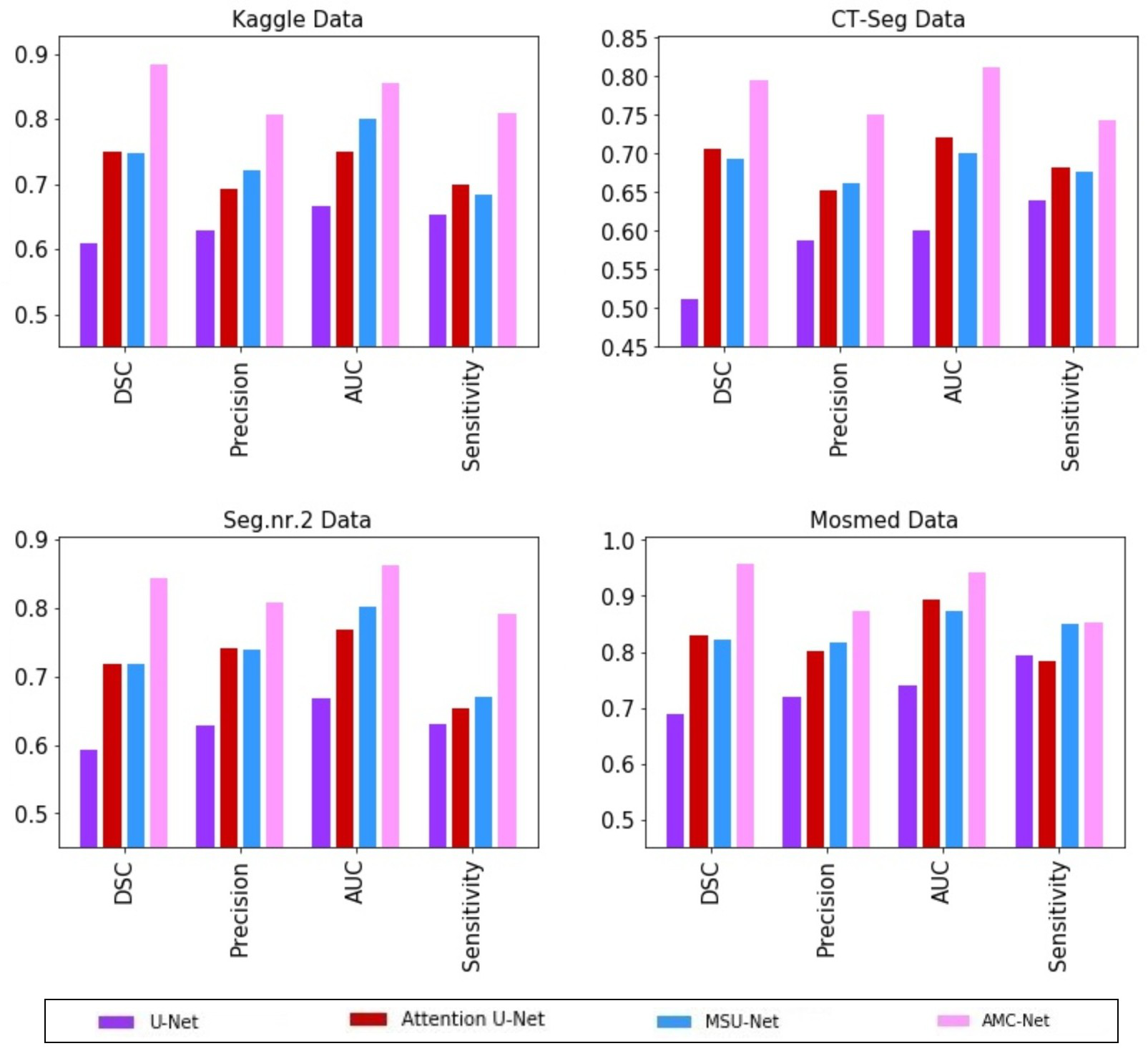}
    \caption{Role of different components of $AMC$-Net, as base classifier in $EAMC$, over the test datasets}
    \label{fig:barPlotAblation}
\end{figure}

The first set of experiments were performed by evaluating the role of each of the components in $AMC$-Net, {\it viz.} $MS$-block  and $AG$, when  used in $EAMC$.  The traditional $U$-Net is thus the baseline, with Attention $U$-Net incorporating only the $AG$. The $U$-Net with only the $MS$-block is termed the $MSU$-Net. The $AMC$-Net is the $U$-Net with both $AG$ and $MS$-block. All four models were trained using the same ensembled $LOPO$ framework. Comparative results on the four test datasets of Table~\ref{data}, evaluated in terms of the performance metrics of eqns. (\ref{Metric:DiceScore})-(\ref{Metric:Sensitivity}), and Area Under the $ROC$ Curve ($AUC$), are presented  in Fig.~\ref{fig:barPlotAblation}. Here It is observed that the proposed $AMC$-Net performs the best, over all the metrics in all test datasets, as compared to the rest (lacking one or more of its modules).   This helps justify the effectiveness of the proposed $EAMC$, which ensembles with LOPO a set of $AMC$-Net models.

    The second task was to explore the effect of the different loss functions of eqns. (\ref{focalLoss})-(\ref{focalTverskyLoss}) on the performance of the base model $AMC$-Net, without any ensembling. Results are provided in Table  \ref{tab:lossfcns}, over all the four test datasets. Here the ROI (GGO and consolidation) covers a minuscule portion of a CT slice. The Focal loss $FL$ is found to be the best because of its capability in handling class imbalance; thereby, reducing misclassification error. As it predicts the outcome as probability, it can better distinguish between grades of severity in outcome. A mechanism of down-weighting the easier samples while emphasizing the more challenging ones, helps $FL$ focus on the smaller ROIs while suppressing the background regions.

    Note that ensembling in $EAMC$ enhances the corresponding performance (as reported in Fig. \ref{fig:barPlotAblation}) in terms of $DSC$.

\begin{table}[htb]
\caption{Effect of various loss functions on $AMC$-Net, measured in terms of $DSC$, over the test datasets}
\centering
\scalebox{0.95}{
\begin{tabular}{@{}lllll@{}}
    \toprule
    Loss function & \begin{tabular}{@{}l@{}}Kaggle-COVID-19: \\ Part-2\end{tabular} &	CT-Seg & Seg-nr.2 & MosMed \\ [0.5ex]
    \midrule
    $DL$  &  0.7521 ± 0.174 & 0.7238 ± 0.201 & 0.7218 ± 0.136 & 0.8465 ± 0.251 \\
    $CEDL$ & 0.7272 ± 0.142 & 0.7377 ± 0.193 & 0.7996 ± 0.147 & 0.8129 ± 0.180 \\
    $IoU$  & 0.7937 ± 0.137 & 0.7190 ± 0.165 & 0.7660 ± 0.125 & 0.8706 ± 0.197 \\
    $FL$ & \textbf{0.8706 ± 0.129} & \textbf{0.7821 ± 0.151} & \textbf{0.8315 ± 0.132} & \textbf{0.9513 ± 0.096} \\
    $TL$  & 0.7103 ± 0.176 & 0.7201 ± 0.130 & 0.7542 ± 0.203 & 0.8360 ± 0.071 \\
    $FTL$  & 0.8238 ± 0.192 & 0.7725 ± 0.144 & 0.8285 ± 0.194 & 0.8973 ± 0.214 \\
    \bottomrule
\end{tabular}
}
\label{tab:lossfcns}
\end{table}

Finally investigations were pursued with different Dilation rates in the convolution layers of the $MS$-block.  It was observed that the combination $D$ = 1, 2, 3, and 5, provides the best results over the test data, in terms of average $DSC$.

\begin{table}[ht]
\caption{Effect of varying Dilation rate $D$ on $AMC$-Net, measured in terms of $DSC$ over the test datasets}
\centering
\scalebox{0.95}{
\begin{tabular}{@{}lllll@{}}
    \toprule
    \begin{tabular}{@{}l@{}}$D$ in $MS$-block\end{tabular}  & \begin{tabular}{@{}l@{}}Kaggle-COVID-19: \\ Part-2\end{tabular} &	CT-Seg & Seg-nr.2 & MosMed \\ [0.5ex]
    \midrule
    1,2,3,4 & 0.8017 ± 0.182 & 0.7018 ± 0.176 & 0.7552 ± 0.150 & 0.8910 ± 0.101 \\
    1,2,4,8 & 0.7482 ± 0.149 & 0.6617 ± 0.168 & 0.7001 ± 0.152 & 0.8527 ± 0.107 \\
    1,2,3,5 & \textbf{0.8706 ± 0.129} & \textbf{0.7821 ± 0.151} & \textbf{0.8315 ± 0.132} & \textbf{0.9513 ± 0.096} \\
    \bottomrule
\end{tabular}
}
\label{tab:DilationChange}
\end{table}

\subsection{Severity assessment}\label{severity}

\begin{table}[!ht]
        \centering
        \caption{Prediction of affected region,  by $EAMC$, considering two samples slices from each test dataset}
          \begin{tabular}{@{}llll@{}}
            \toprule
            Dataset & \begin{tabular}{@{}l@{}}Sample no. \\ in Fig. ~\ref{fig:segmentation}\end{tabular} &	\begin{tabular}{@{}l@{}}Ground \\ truth (\%) \end{tabular} &	\begin{tabular}{@{}l@{}} Prediction  \\ (\%) by $EAMC$ \end{tabular} \\
		      \midrule

            \multirow{2}{*}{\begin{tabular}{@{}l@{}}Kaggle-COVID-19: \\ Part-2\end{tabular}}		
		& S1 & 57.63 & 57.71 \\
            & S2 &  85.28 & 84.71 \\
            \midrule

            \multirow{2}{*}{CT-Seg}
            & S3 & 89.11 & 88.12 \\
	    & S4 & 78.82 & 79.34 \\
            \midrule

            \multirow{2}{*}{Seg-nr.2}
	    & S5 & 60.91 & 61.15 \\
	    & S6 & 62.71 & 62.18 \\
            \midrule

            \multirow{2}{*}{MosMed}
	    & S7 & 11.62 & 10.98 \\
	    & S8 & 13.74 & 14.02 \\
            \bottomrule
        \end{tabular}
        \label{tab:Severity}
    \end{table}

The methodology of grading the severity of COVID-19 infection, as developed by the  Russian Federation \cite{morozov2020Mosmeddata2}, is based on individually computing the volume ratio of lesions in each lung and using the maximal value to assess the overall severity score. The  range of the score is divided into five categories based on volume of damaged lung tissue. \begin{itemize} \item CT-0: not consistent with pneumonia (including COVID-19), $ie$, normal \item CT-1: infection involvement of $\le$ 25 \% \item CT-2: infection involvement of 25-50 \%  \item CT-3: infection involvement of 50-75 \% \item CT-4: infection involvement of 75-100 \% \end{itemize} Patients with CT-3 (severe pneumonia) or higher are typically hospitalized, with  CT-4 (acute pneumonia) required to be admitted to an intensive care unit.

Table \ref{tab:Severity} quantifies the infected region in the sample test images, comprising  the eight slices (two each from the four sets of test data) of Fig~\ref{fig:segmentation}, along similar lines.

\begin{figure}[!ht]
    \centering
    \includegraphics[width=16cm]{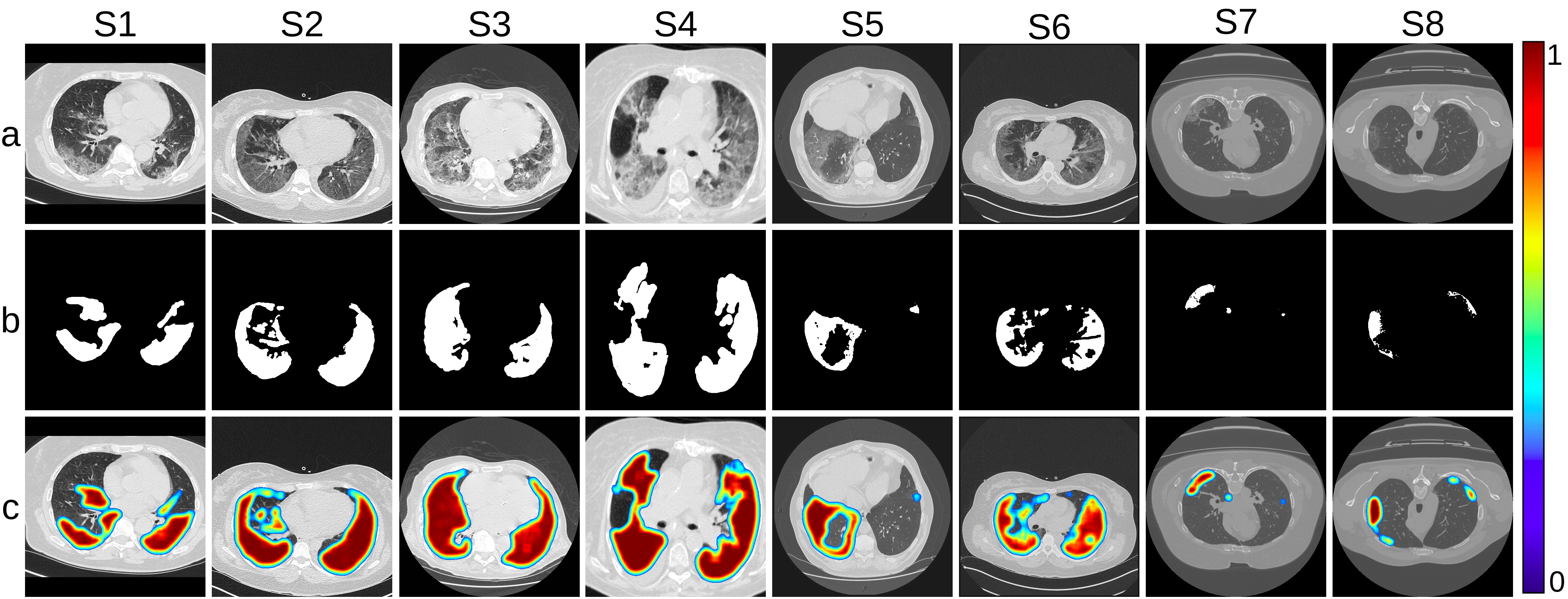}
    \caption{Visualization of infection, in the eight sample slices from Fig. \ref{fig:segmentation}, as predicted by $EAMC$. (a) Input CT scan, (b) corresponding annotation, and (c) JET ColorMap on the prediction}
    \label{fig:color}
\end{figure}

Fig.~\ref{fig:color} depicts the qualitative assessment of the severity of infection, over the same eight sample slices, based  on the color mask of the segmented output generated by JET Colormap \cite{wang2016JET}.
The visualization uses the predicted  probability values of the various regions, to illustrate the grading of severity. Here red corresponds to the  most severe, yellow indicates moderate, and blue refers to the least severe infections. Such analysis can be of assistance in predicting the grading of infection in a sample patient, along with an estimation of the expected prognosis.

\section{Discussion}\label{Conclusions}

Medical imaging provides useful assistance for the safe, efficient, and early detection, diagnosis, isolation, and prognosis of diseases. It enables non-invasive examination of the interior organs, bones and tissues, allowing for accurate assessment of disease severity. Particularly, with the advancement in CT imaging technology, very high resolution images serve as  suitable diagnostic tool in the medical domain. The recent COVID-19 pandemic demonstrated that CT images are often more accurate as compared to the standard RT-PCR tests, which can exhibit False Positives.  The CT images can provide a lot more information, like the severity of infection and the presence and distribution of pathologies like GGO and Consolidation.  Therefore, CT images are gradually becoming a primary tool for the detection and prognosis in COVID-19. As the increased number of cases for diagnosis made the job over-burdening, the need for automated and more accurate segmentation, detection and analysis became evident. 

Our research using deep convolutional networks in medical imaging analysis, demonstrated efficient  extraction of valuable features. The results depict how observable features from the COVID-19 ROI, encompassing GGOs and consolidations, could be effectively retrieved from various blocks at different levels. The severity of the disease could also be assessed. The qualitative and quantitative statistics illustrate the superiority of our model with respect to related methods in literature. The qualitative output demonstrates that the proposed $EMAC$ generates very few misclassified pixels corresponding to the ROI.  
The data revealed the presence of a substantial proportion of pixels from the background region, with a relatively smaller number from the ROI. Such a significant class imbalance is effectively addressed by the loss function discussed. Our patch-based method performs exceptionally well, in terms of accuracy and loss over training and validation, for an effective management of overfitting in deep learning. Use of CT images, obtained from several other sources, established the robustness of our methodology in handling imaging differences at the source.

A novel ensembling method by $LOPO$ was developed for a collective, efficient delineation of COVID-19 affected region in the lung, along with a gradation of the severity of the disease, using very limited training data.  Multi-scalar attention with deep supervision enabled enhanced accuracy, in terms of improved sensitivity and precision in segmentation of ROI, for the proposed model $EAMC$. The loss function helped focus on the imbalanced representation of the ROIs, in terms of GGOs and consolidations in the CT slices. While the training was performed on one set of annotated data, the testing set comprised of an assortment of data from different sources of publicly available sets. The superiority of the network was thus established in a broader generalization framework.

 \begin{figure}[!ht]
    \centering
    \includegraphics[width=16cm, height=15cm]{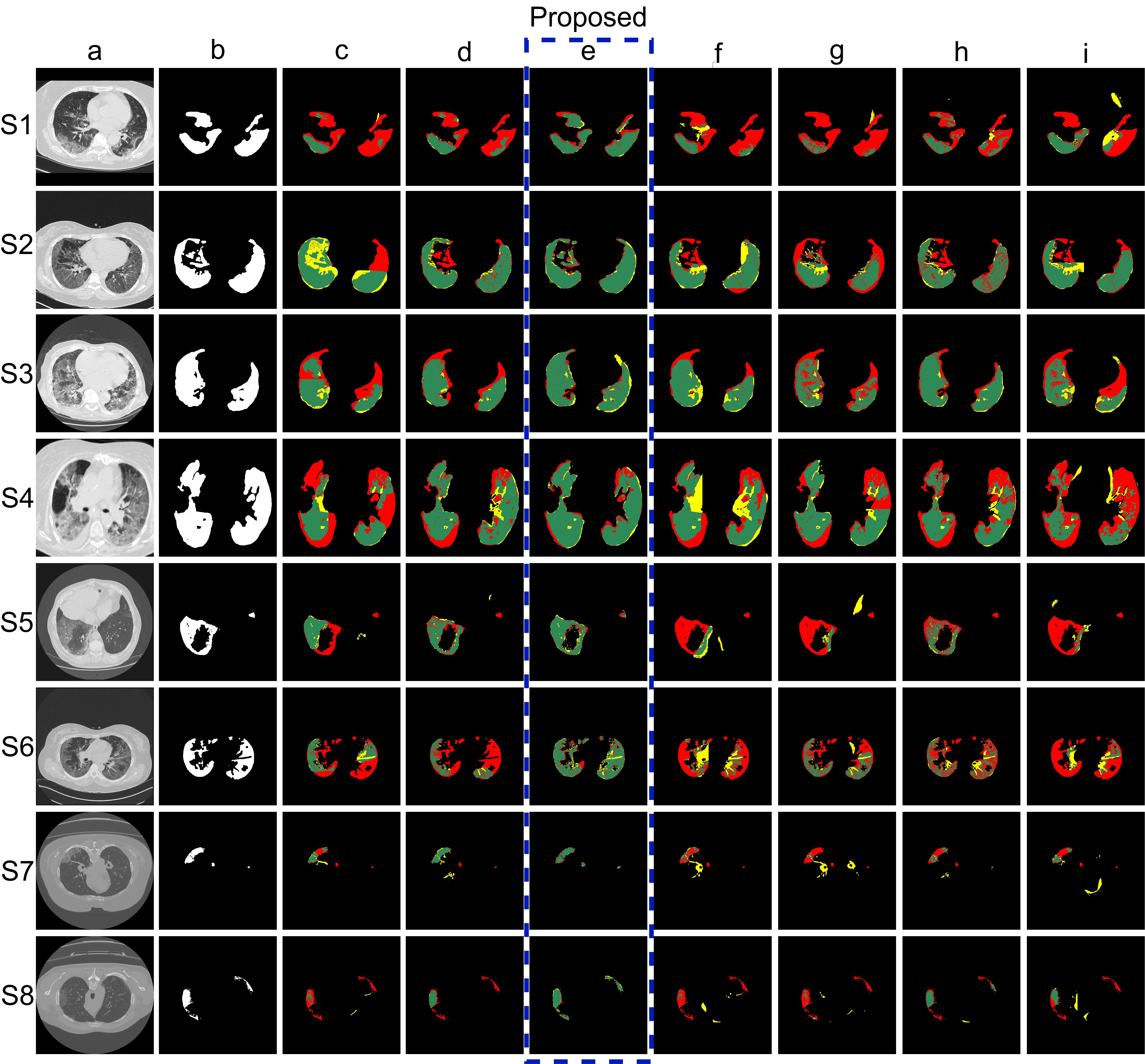}
    \caption{Sample segmentation output by base classifiers, in the uniform framework of ensembling with $LOPO$. (a) Original CT scan, with (b) Annotated masks. Segmentation obtained by (c) $U$-Net, (d) Attention $U$-Net, (e) $AMC$-Net, (f) R2UNet, (g) Inception Net, (h) Xception Net, and (i) DenseNet, where regions Green: True Positive, Red: False Negative, Yellow: False Positive}
    \label{fig:segmentation}
\end{figure}

\begin{figure}[!ht]
    \centering
    \includegraphics[width=15cm]{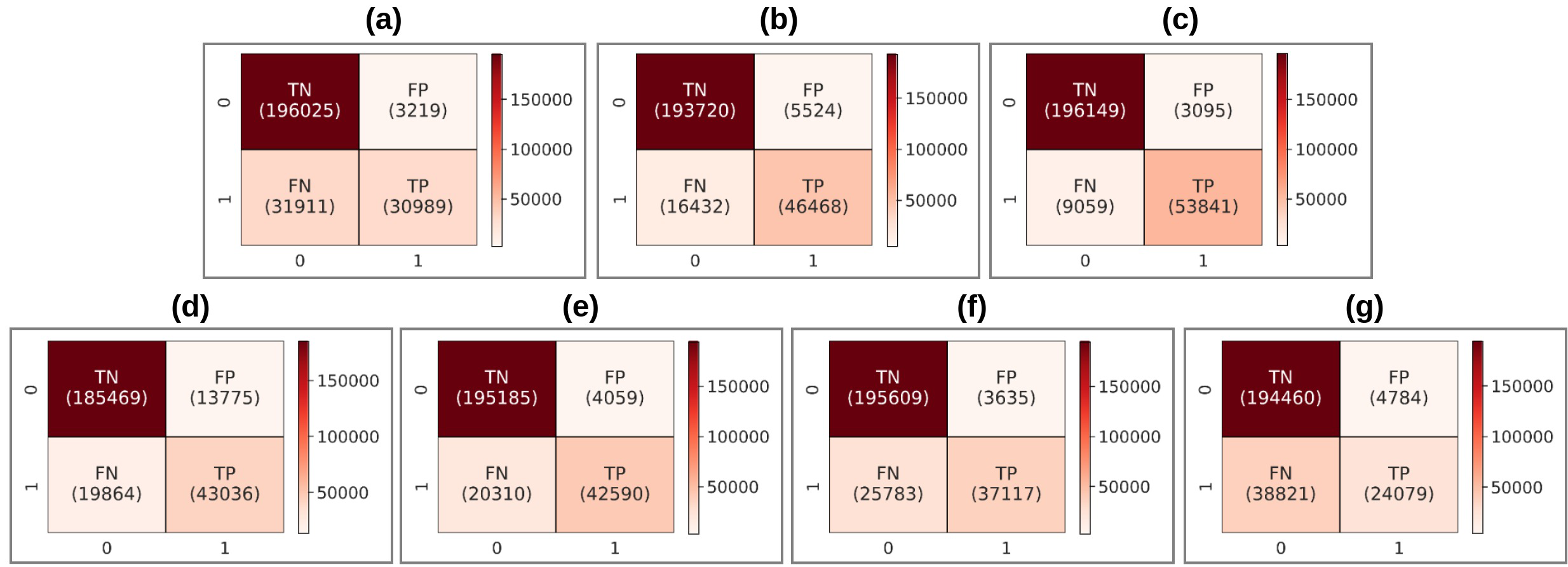}
    \caption{Confusion matrix for a sample $S4$ from Fig. \ref{fig:segmentation} segmentation mask, by the base classifiers, in the uniform framework of ensembling with $LOPO$. Models (a) $U$-Net, (b) Attention $U$-Net, (c) $AMC$-Net, (d) R2UNet, (e) Inception Net, (f) Xception Net, and (g) DenseNet}
    \label{fig:confusionMatrix}
\end{figure}

Sample qualitative results in  Fig.~\ref{fig:segmentation}, corresponding to the models compared in Fig.~\ref{fig:boxplotComparison}, help establish the robustness and effectiveness of $EAMC$ under ensembling by $LOPO$ on the backbone $AMC$-Nets. The output is evaluated in terms of the annotated masks. The eight samples explored were (i) $S1$, $S2$ from Kaggle-COVID-19: Part-2; (ii) $S3$, $S4$ from CT-Seg; (iii) $S5$, $S6$ from Seg-nr.2; and (iv) $S7$, $S8$ from MosMed. It is observed that the $AMC$-Net, of column (e) in the figure, performed the best. Results were corroborated with the confusion matrix of Fig.~\ref{fig:confusionMatrix}, providing an indication of the distribution of misclassified pixels for a sample segmentation mask $S4$. The confusion was seen to be the least in case of the $AMC$-Net [column (c)]. It resulted in the minimum over-and under-segmentation for all eight samples considered here.

Note that over-segmentation corresponds to higher count of $FP$ pixels (yellow) and under-segmentation refers to higher $FN$ (red). Considering sample $S4$ as an example,  it is clearly evident that state-of-the-art models $U$-Net, Attention $U$-Net, Inception Net, Xception Net, DenseNet demonstrate under-segmentation, while model R2UNet is indicative of over-segmentation.  On the other hand, our $AMC$-Net [column (e), Fig. \ref{fig:boxplotComparison}] exhibited significantly lower under- and/or over-segmentation for the same sample $S4$. The corresponding confusion matrix in Fig. \ref{fig:confusionMatrix}  corroborates these findings.

 \begin{table}[!ht]
 \centering
 \caption{Comparative computational analysis of base model parameters, along with $DSC$ obtained}
 \scalebox{0.8}{
\begin{tabular}{@{}llcccccccc@{}}
 \toprule
& Model & $U$-Net & \begin{tabular}{@{}c@{}}Attention \\ $U$-Net \end{tabular} & $MSU$-Net & $AMC$-Net & R2UNet & \begin{tabular}{@{}c@{}}Inception \\ Net \end{tabular} & \begin{tabular}{@{}c@{}}Xception \\ Net \end{tabular} & DenseNet\\
\midrule
& \#parameters & 1.96M & 1.99M & 3.31M & 3.34M & 6.00M & 11.99M &  2.05M & 4.26M \\
\midrule
\parbox[t]{2mm}{\multirow{4}{*}{\rotatebox[origin=c]{90}{\ul{ DSC on }}}}
& Kaggle & 0.6082 & 0.7492 & 0.7477 & \textbf{0.8840} & 0.6772 & 0.5924 & 0.7972 & 0.6729 \\
& CT-Seg & 0.5124 & 0.7054 & 0.6924 & \textbf{0.7955} & 0.5762 & 0.6327 & 0.7172 & 0.6766  \\
& Seg-nr.2 & 0.5921 &	0.7180 & 0.7173 & \textbf{0.8431} & 0.6304 & 0.6275 & 0.8331 & 0.8105  \\
& MosMed & 0.6896 & 0.8300 & 0.8224 & \textbf{0.9584} & 0.7029 & 0.7145 & 0.8908 & 0.8220 \\
\bottomrule
\label{tab:complex}
\end{tabular}
}
\end{table}

The model complexity, in terms of the number of parameters, is enumerated in  Table~\ref{tab:complex}. Although the proposed $EAMC$ involves slightly more parameters than $U$-Net, Attention $U$-Net, $MSU$-Net, Xception Net, it is less than that of R2UNet, DenseNet, and much lower as compared to Inception Net. Note that the overall comparative performance of $EAMC$ is better than the state-of-the-art methods, as evident in Fig. \ref{fig:boxplotComparison}. A representative study for $DSC$ on all the four test datasets is also provided in the table.

The technique holds promise in other medical image domains, including (but not limited to) detection of lesions in MRI images of the brain or pathologies in fundus images of the eye for screening diabetic retinopathy.

\section{Methodology}\label{Methodology}

 The architecture of our $EAMC$ is novel. It consists of an ensemble of Attention-modulated Multi-Scalar ($MS$) blocks along the encoding and decoding paths of a U-Net. Incorporation of dilated convolutions in the $MS$-block, in lieu of down- and/or up-sampling, improves the overall performance and makes it robust to generalization. Focal loss function \cite{lin2017focal} is  employed for effectively handling class imbalance in the data. The Leave-One-Patient-Out (LOPO) ensembling effectively trains a network from scratch (each time leaving one patient sample out). This scheme creates the necessary diversity in data, while training a completely new model with different parameters and scarce annotated samples.

\begin{figure}[!ht]
    \centering
    \includegraphics[width=16cm]{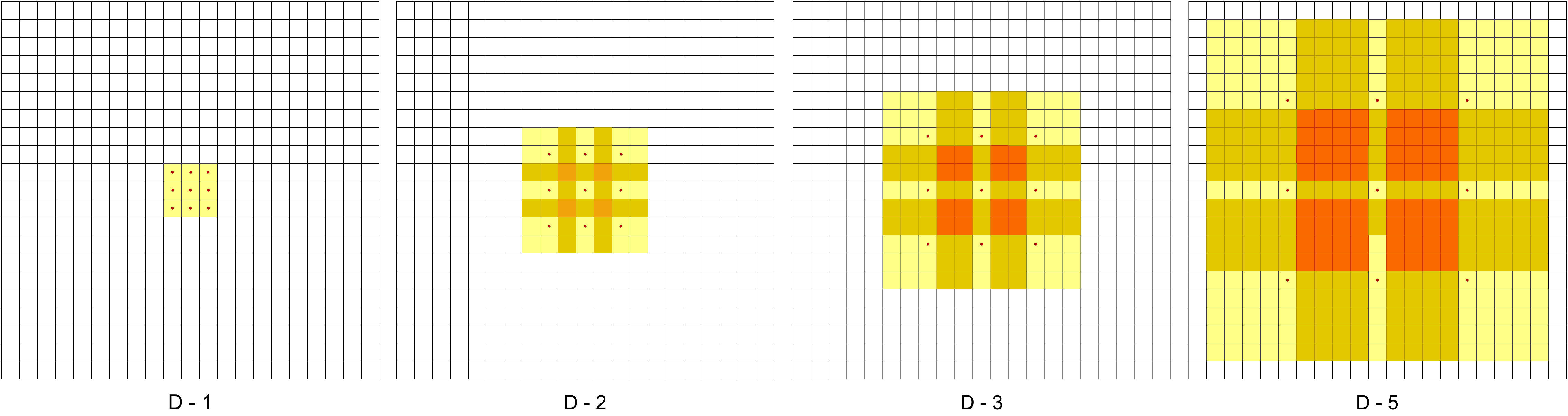}
    \caption{$D$-dilated convolutions, with $D =  1, 2, 3, 5$}
    \label{fig:Dilation}
\end{figure}

The Attention-modulated $MS$-blocks form the  {\it AMC-Net} modules, serving as the base classifier for our ensembled $EAMC$.  This is illustrated in Fig.~\ref{fig:AMCmodel}, with elaborated description of the individual modules being provided in Tables \ref{AMC-Net Architecture}-\ref{AG-block}. The $MS$-block extracts multi-scalar features using a concatenation of four dilated convolutional layers, having dilation rates  $D =  1, 2, 3, 5$.
A dilated convolution (Fig.~\ref{fig:Dilation}) inserts holes into the standard convolution map, thereby expanding its receptive fields. Thus  dilated convolutions can enlarge a receptive field,  without any loss of information, while retaining the kernel size. Finally $1 \times 1$ convolutions are employed on the concatenated multi-scalar feature map. The $MS$-block is depicted in Fig.~\ref{fig:AMCmodel}(b) and Table~\ref{tab:MS-Block}.

\begin{table}[!ht]
\begin{minipage}[c]{0.5\textwidth}
\centering
\caption{{\it AMC-Net} module of Fig.~\ref{fig:AMCmodel}(a)}
\scalebox{0.7}{
\begin{tabular}{@{}ll@{}}
    \toprule
    Block: i & Input: (128 * 128 * 1)  CT image\\
    [0.5ex]
    \midrule

    \multirow{4}{*}{1} & conv$_{1}$(3*3), 16 \\
    & \textcolor{cyan}{MS-Block}(conv$_{1}$), 16 (Described in Table~\ref{tab:MS-Block})\\
    & conv$_{2}$(3*3), 16 \\
    & maxpooling$_{1}$(2*2) \\
    \midrule

    \multirow{4}{*}{2} & conv$_{3}$(3*3), 32 \\
    & \textcolor{cyan}{MS-Block}(conv$_{3}$), 32 \\
    & conv$_{4}$(3*3), 32 \\
    & maxpooling$_{2}$(2*2) \\
    \midrule

    \multirow{4}{*}{3} & conv$_{5}$(3*3), 64 \\
    & \textcolor{cyan}{MS-Block}(conv$_{5}$), 64 \\
    & conv$_{6}$(3*3), 64 \\
    & maxpooling$_{3}$(2*2) \\
    \midrule

    \multirow{4}{*}{4} & conv$_{7}$(3*3), 128 \\
    & \textcolor{cyan}{MS-Block}(conv$_{7}$), 128 \\
    & conv$_{8}$(3*3), 128 \\
    & maxpooling$_{4}$(2*2) \\
    \midrule
    \midrule
    \multirow{4}{*}{5} & conv$_{9}$(3*3), 256 \\
    & conv$_{10}$,(3*3), 256 \\
    & Upsampling$_{1}$(2*2) \\
    & AG(conv$_{8}$, Upsampling$_{1}$), 64 (Described in Table~\ref{AG-block})\\
    \midrule
    \midrule
    \multirow{6}{*}{6} & concat$_{1}$ (Upsampling$_{1}$, Multiply$_{1}$) \\
    & conv$_{11}$(3*3), 128 \\
    & \textcolor{cyan}{MS-Block}(conv$_{11}$), 128 \\
    & conv$_{12}$(3*3), 128 \\
    & Upsampling$_{2}$(2*2) \\
    & AG (conv$_{6}$ Upsampling$_{2}$), 32 \\
    \midrule

    \multirow{6}{*}{7} & concat$_{2}$ (Upsampling$_{2}$, Multiply$_{2}$) \\
    & conv$_{13}$(3*3), 64 \\
    & \textcolor{cyan}{MS-Block}(conv$_{13}$), 64 \\
    & conv$_{14}$(3*3), 64 \\
    & Upsampling$_{3}$(2*2) \\
    & AG (conv$_{4}$, Upsampling$_{3}$), 16 \\
    \midrule

    \multirow{6}{*}{8} & concat$_{3}$ (Upsampling$_{3}$, Multiply$_{3}$) \\
    & conv$_{15}$(3*3), 32 \\
    & \textcolor{cyan}{MS-Block}(conv$_{15}$), 32 \\
    & conv$_{16}$(3*3), 32 \\
    & Upsampling$_{4}$(2*2) \\
    & AG (conv$_{2}$, Upsampling$_{4}$), 8 \\
    \midrule

    \multirow{6}{*}{9} & concat$_{4}$ (Upsampling$_{4}$, Multiply$_{4}$) \\
    & conv$_{17}$(3*3), 16 \\
    & \textcolor{cyan}{MS-Block}(conv$_{17}$), 16 \\
    & conv$_{18}$(3*3), 16 \\

    & conv$_{19}$(1*1), 1 \\
    & Sigmoid activation \\
    \midrule
    & Output: (128 * 128 * 1) \\
    \bottomrule
\end{tabular}
\label{AMC-Net Architecture}
}
\end{minipage}
\begin{minipage}[c]{0.5\textwidth}
\centering
\caption{$MS$-block module of Fig.~\ref{fig:AMCmodel}(b)}
\scalebox{0.7}{
\begin{tabular}{@{}l@{}}
    \toprule
        \textcolor{cyan}{MS-Block}(conv$_M$), n (n = No. of features) \\ [0.5ex] 
        \midrule
        Input(conv$_M$) -$>$ conv$_{D-1}$(3*3), n \\
        Input(conv$_M$) -$>$ conv$_{D-2}$(3*3), n \\
        Input(conv$_M$) -$>$ conv$_{D-3}$(3*3), n \\
        Input(conv$_M$) -$>$ conv$_{D-5}$(3*3), n \\
        concat (conv$_{D-1}$, conv$_{D-2}$, conv$_{D-3}$, conv$_{D-5}$) \\
        \bottomrule
    \end{tabular}
    }
    \label{tab:MS-Block}

\vspace*{4 cm}

\caption{Attention Gates ($AG$) of  Fig.~\ref{fig:AMCmodel}(c)}
\scalebox{0.7}{
\begin{tabular}{@{}l@{}}
    \toprule
        AG (conv$_X$, Upsampling$_Y$),  n (n = No. of features) \\ [0.5ex]
        \midrule
        Input(Upsampling$_Y$) -$>$ conv$_a$(1*1), n \\
        Input(conv$_X$) -$>$ conv$_b$(1*1), n \\
        conv$_a$ + conv$_b$ \\
        ReLU activation \\
        conv$_c$(1*1), 1 \\
        Sigmoid Activation -$>$ SA \\
        conv$_X$ * SA -$>$ Multiply$_Y$ \\
        \bottomrule
    \end{tabular}
    }
    \label{AG-block}

\end{minipage}
\end{table}

The {\it AMC-Net} contains nine convolutional blocks, four max-pooling layers, four Upsampling
convolutional layers and eight $MS$-blocks. First the CT image patches of  size  $128 \times  128$ are fed at the input. The patches percolate down four sets of iterations of $2 \times 2$ max-pooling layers, $3 \times 3$ convolutional and $MS$-block layers, involving a stride of 1 in the encoder. The $2 \times 2$ Upsampling layers  in the decoder help recover the final resolution of an image.

The $MS$-blocks are added after the ordinary convolutions in the first four encoder and the last four decoder layers. They help obtain multi-scalar contextual information, to reduce the error along the segmentation boundary for improved accuracy. The COVID-19 infection lesions are typically hard to segment; mainly due to their uneven distribution and varying dimensions. The high-level semantic feature maps in the decoder, concatenated through the attention mechanism, focus on the low-level details in the extracted  feature maps (of the encoder) to accurately recover the details of the infected regions in the CT slices.

The Attention gates ($AG$) of Fig.~\ref{fig:AMCmodel}(c) provide the necessary importance to each pixel during  decoding. The upsampled images, along with their encoded versions at the same level, are combined to enhance the importance of a pixel through spatial attention.
Adaptive selection of spatial information is achieved by emphasizing pixels from the regions of interest, while suppressing the less relevant ones. Four attention modules are  introduced for adaptive feature refinement. A sequential spatial attention module is embedded into  each decoding block to avoid overfitting, while accelerating the training of the $EAMC$.

The activation function at the final layer of the {\it AMC-Net} is the sigmoid. It generates a probabilistic output for the ROI. The choice of a loss function has direct impact on model performance. Loss functions represent the computation of error over each batch, during backpropagation training, and reflect the adjustment of network weights. It was found, after several experiments, that  focal loss was the best choice.  Let the ground truth segmentation mask be $\mathbf{y} \in$ \{±1\}, with the corresponding predicted mask being $\mathbf{\hat{y}}$ with estimated probability $p \in$ [0,1]. Focal Loss ($FL$)~\cite{lin2017focal} overcomes class imbalance in datasets, where positive patches are relatively scarce. The cross entropy ($CE$) loss for binary classification is defined as
\begin{equation}
\label{CE}
  CE(p,y)=\begin{cases}
    -\log{p}, & \text{if $y=1$},\\
    -\log(1-p), & \text{otherwise}.
  \end{cases}
\end{equation}
For convenience, let
\begin{equation}
p\textsubscript{t} = \begin{cases}
    p, & \text{if $y=1$},\\
    1-p, & \text{otherwise},
    \end{cases}
\end{equation}
such that $CE(p,y) = CE(p\textsubscript{t}) = -log(p\textsubscript{t})$.
The $\alpha$-balanced focal loss is defined as
\begin{equation}
\label{focalLoss}
FL(p\textsubscript{t}) = - \alpha(1 - p_t)^\gamma log(p_t),
\end{equation}
with the choice of weighting factor $\alpha = 0.8$ and focusing parameter $\gamma = 2$ being made after several experiments.

Some of the other loss functions, explored in the ablation studies, include the Dice Loss ($DL$) \cite{DICE-Zhang-2021}
\begin{equation}
\label{diceLoss}
DL(y, \hat{y}) = \biggl(1 - \frac{2y\hat{y} + 1}{y + \hat{y} + 1}\biggr) \,,
\end{equation}
where 1 is added in the numerator and denominator to ensure that in edge case scenarios, such as when $\mathbf{y}$ = $\mathbf{\hat{y}}$ = 0, the function does not become undefined. The $CEDL$ loss \cite{lin2017focal} is defined as a combination of $DL$ and the cross-entropy ($CE$) loss for binary classification, thereby incorporating the benefits from both. We have
\begin{equation}
\label{CEDiceLoss}
CEDL(y,\hat{y}) = - \{y\log_{}(\hat{y}) + (1 - y)\log_{}(1 - \hat{y})) + DL(y,\hat{y})\}.
\end{equation}
The $IoU$ metric \cite{IoU-Zhou-2019}, or Jaccard Index, is computed as the ratio between the overlap of the positive instances between two sets, and their mutual combined values. It is expressed as
\begin{equation}
\label{IoULoss}
IoU(y,\hat{y}) = \biggl(1 - \frac{y\hat{y} + 1}{y + \hat{y} + y\hat{y}+ 1}\biggr) \,.
\end{equation}
The Tversky loss ($TL$) \cite{Tversky-Abraham-2019} optimises the segmentation on imbalanced medical datasets. It adjusts the constants $\alpha$ and $\beta$ to give special weightage to errors like $FP$ and $FN$  We have
\begin{equation}
\label{tverskyLoss}
TL(y,\hat{y}) = \biggl(1 - \frac{y\hat{y} + 1}{y\hat{y} + \beta(1-y)\hat{y} + \alpha y(1 - \hat{y}) + 1}\biggr) \,.
\end{equation}
The Focal Tversky loss ($FTL$) \cite{Tversky-Abraham-2019} also focuses on the difficult samples, by down-weighting easier (or common) ones. It attempts to learn the harder examples, like small ROIs, with the help of the $\gamma$ coefficient. It is defined as
\begin{equation}
\label{focalTverskyLoss}
FTL(y,\hat{y}) = (1 - TL)^{1/\gamma}.
\end{equation}
A value of $\gamma = 2$ was employed, after several experiments.

Diversity is introduced in this ensembling of the {\it AMC-Nets}, by varying the training datasets through $LOPO$; thereby, changing the initialization of the networks, and modulating the choice of  parameters of the $EAMC$ system.
The performance of the models is evaluated in terms of the following metrics. We define the number of pixels, (i) correctly classified as positive by True Positive ($TP$), (ii) incorrectly classified as positive, by False Positive ($FP$), (iii) correctly identified as negative as True Negative ($TN$), and falsely classified as negative by False Negative ($FN$). \\ The metrics used are {\it Dice Score Coefficient}
\begin{equation}
\label{Metric:DiceScore}
    \mathbf{DSC} = \biggl(\frac{2*TP}{2*TP + FP + FN}\biggr) \,,
\end{equation}
\begin{equation}
\label{Metric:Precision}
    \mathbf{Precision} = \biggl(\frac{TP}{TP+FP}\biggr) \,,
\end{equation}
\begin{equation}
\label{Metric:Sensitivity}
    \mathbf{Sensitivity} = \biggl(\frac{TP}{TP+FN}\biggr) \,,
\end{equation}
and Area Under the {\it Receiver Operating Characteristic} Curve ($AUC$). The $ROC$ curve typically plots  the $TP$ rate vs the $FP$ rate, over different thresholds. Higher values of these indices imply a better quality of segmentation \cite{song2022covid,wang2022regularization}.

\section{Data Preparation}\label{datasets}

Pixel values in range [0, 255] were normalized, keeping the HU range in interval [-1024, 3071], to enable the model visualize and learn all the areas (like, infection, bone, tissues) inside the CT scan images. Instead of initially extracting the lung part from the full CT slice \cite{wu2020severity}, we directly detect the infected area from the entire image for subsequent segmentation of the COVID-infected ROI. Class imbalance between the infected and non-infected areas of the CT slices was considered, in terms of positive (infected) and negative (non-infected) patches over the data.

\begin{figure}[!htb]
    \centering
    \includegraphics[width=10cm]{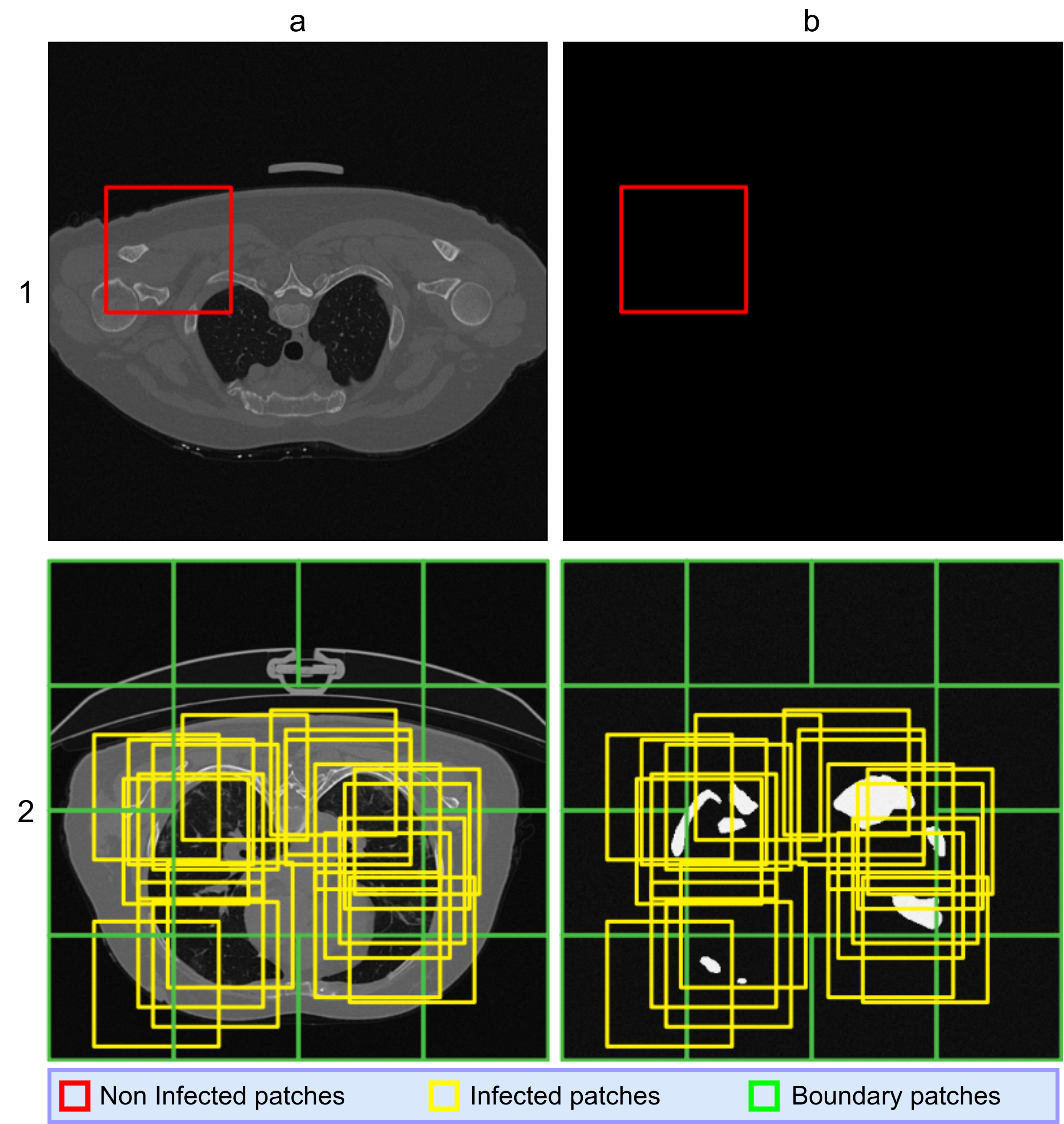}
    \caption{\emph{Row 1:} Patch extraction from training CT slices, depicting no infected regions. 1(a) Non-infected patches, and 1(b) corresponding annotated masks.\\ \emph{Row 2:} Patch extraction from training CT slices, depicting infected regions. 2(a) Overlapping patches, and 2(b) mapping to corresponding annotated masks.}
    \label{fig:fig3}
\end{figure}

\subsection{Training} \label{TrainPatch}

Availability of annotated training data, depicting infection masks, is scarce and leads to class imbalance. In order to circumvent this problem, we extracted  overlapping patches to increase the training data while uniformly representing relevant ROI.

The ground truth corresponding to each axial slice, of each CT volume of the training data was checked. If there existed no infected region on a slice then it was labeled as ``non-infected" (Fig. \ref{fig:fig3}, \emph{Row: 1}). Random $128 \times 128$ patches were extracted.

When there existed a region of infection in any axial slice, it was labeled as ``ìnfected" (Fig. \ref{fig:fig3},  \emph{Row: 2}). Twenty random $128 \times 128$ bounding boxes were drawn over the ROI to extract the patches. Next
all twelve $128 \times 128$ boundary patches (inside the $512 \times 512$ axial slice) were considered.

 \begin{table}[!ht]
        \centering
        \caption{Distribution of infected and non-infected Slices \& Patches extracted for training}
        \scalebox{1}{
        \begin{tabular}{@{}llcccc@{}}
            \toprule
             \begin{tabular}{@{}l@{}}Patient \\ No.\end{tabular}  & Sample Name & \multicolumn{2}{c}{Slices} & \multicolumn{2}{c}{Patches}\\
             & & Infected & Non-infected & Infected & Non-infected \\
            \midrule
             P1 & coronacases\_org\_001	& 161 &	140 &	3534 &	1758\\
            P2 & coronacases\_org\_002 &	143 &	57 &	3114 &	1519\\
            P3 & coronacases\_org\_003 &	137 &	63 &	3198 &	1249\\
            P4 & coronacases\_org\_004 &	113 &	157 &	2341 &	1432\\
            P5 & coronacases\_org\_005 &	116 &	174 &	2342 &	1544\\
            P6 & coronacases\_org\_006 &	70 &	143 &	1503 &	880\\
            P7 & coronacases\_org\_007 &	93 &	156 &	2067 &	1065\\
            P8 & coronacases\_org\_008 &	216 &	85 &	4647 &	2350\\
            P9 & coronacases\_org\_009 &	111 &	145 &	2276 &	1421\\
            P10 & coronacases\_org\_010 &	191 &	110 &	4375 &	1847\\
            \bottomrule
        \end{tabular}
        }
        \label{tab:patchDataSampleBased}
    \end{table}

\begin{figure}[!ht]
    \centering
    \includegraphics[width=16cm]{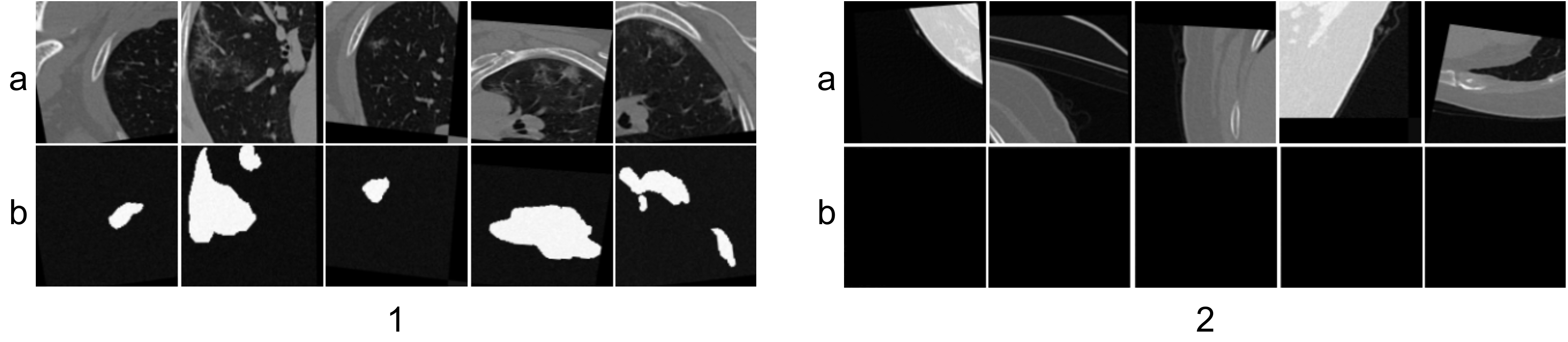}
    \caption{The (a) extracted patches, and (b) corresponding annotation (post-run-time augmentation), for sample slices which are 1: infected, and 2: non-infected }
    \label{fig:fig6}
\end{figure}

The distribution of infected and non-infected slices and/or patches, after patch extraction to create the training set, is displayed in Table~\ref{tab:patchDataSampleBased}.
Representative extracted patches, along with the corresponding annotated masks, are presented in Fig.~\ref{fig:fig6}.1 for the infected slices and  Fig.~\ref{fig:fig6}.2 for the non-infected ones.

\subsection{Testing}\label{sec:secTestPatch}

\begin{figure}[!ht]
    \centering
    \includegraphics[width=5cm]{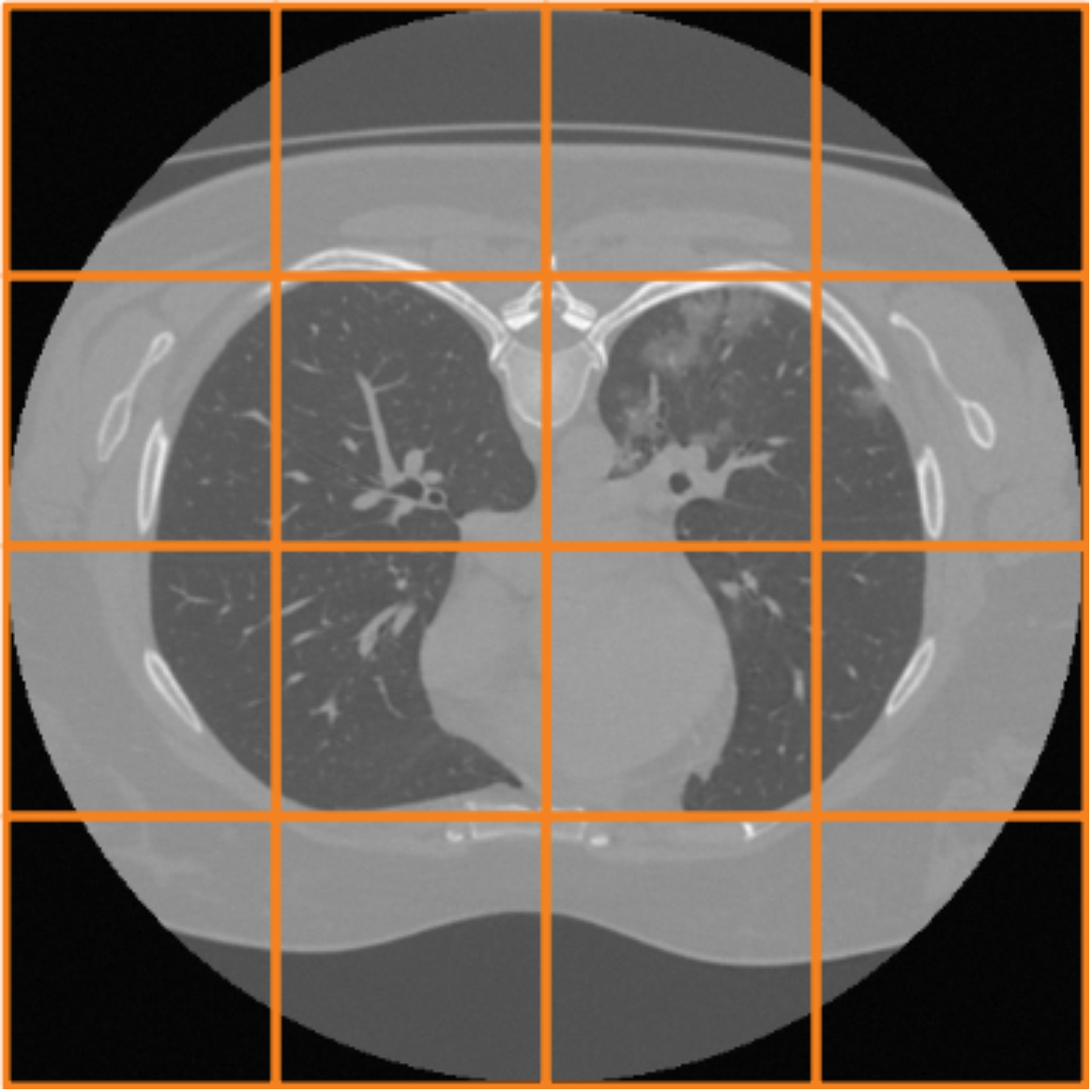}
    \caption{Patch extraction from test CT slices.}   \label{fig:fig8}
\end{figure}

As only the ROI and background need to be separated for the test images, here the extraction of non-overlapping patches serve the purpose. Axial slices ($512 \times 512$) were extracted from each test CT volume. Sixteen $128 \times 128$ non-overlapping patches were obtained from each slice, as depicted in  Fig.~\ref{fig:fig8}.

\bibliographystyle{abbrv}
\bibliography{Sorichi}{}

\end{document}